\documentclass[useAMS,usenatbib]{mn2e}

\usepackage{graphicx}
\usepackage{times}
\usepackage{natbib}
\usepackage{rotating}

\renewcommand{\vec}[1]{\bmath{#1}}
\newcommand{\be}{\begin{equation}}
\newcommand{\ee}{\end{equation}}
\newcommand{\ba}{\begin{eqnarray}}
\newcommand{\ea}{\end{eqnarray}}
\newcommand{\brr}{\begin{array}}
\newcommand{\err}{\end{array}}
\newcommand{\bc}{\begin{center}}
\newcommand{\ec}{\end{center}}

\newcommand{\mincir}{\raise
  -2.truept\hbox{\rlap{\hbox{$\sim$}}\raise5.truept \hbox{$<$}\ }}
\newcommand{\magcir}{\raise
  -2.truept\hbox{\rlap{\hbox{$\sim$}}\raise5.truept \hbox{$>$}\ }}
\newcommand{\siml}{\raise
  -2.truept\hbox{\rlap{\hbox{$\sim$}}\raise5.truept \hbox{$<$}\ }}
\newcommand{\simg}{\raise
  -2.truept\hbox{\rlap{\hbox{$\sim$}}\raise5.truept \hbox{$>$}\ }}

\title[Relative velocity of dark matter and barions in clusters of
galaxies]{Relative velocity of dark matter and barions in clusters of
galaxies and measurements of their peculiar velocities}

\author[K. Dolag, R. Sunyaev]
{K.~Dolag$^{1,2}$\thanks{E-mail: kdolag@mpa-garching.mpg.de}, and R. Sunyaev$^{2,3}$ \\
$^1$ University Observatory Munich, Scheinerstr. 1, 81679 Munich, Germany\\
$^2$ Max-Planck-Institut f\"ur Astrophysik, Karl-Schwarzschild Strasse
  1, Garching bei M\"unchen, Germany\\
$^3$ Space Research Institute (IKI), Russian Academy of Sciences, Profsoyuznaya
str. 84/32, Moscow, 117997 Russia \\
}

\begin{document}

\date{Accepted ???. Received ???; in original form ???}

\pagerange{\pageref{firstpage}--\pageref{lastpage}} \pubyear{0000}

\maketitle

\label{firstpage}

\begin{abstract}
The increasing sensitivity of current experiments, which nowadays routinely measure the
thermal SZ effect within galaxy clusters, provide the hope that 
peculiar velocities of individual clusters of galaxies will be measured rather soon
using the kinematic SZ effect. Also next generation of X-ray telescopes with
microcalorimeters, promise first detections of the motion of 
the intra cluster medium (ICM) within clusters.
We used a large set of cosmological, hydrodynamical simulations, which cover very large 
cosmological volume, hosting a large number of rich clusters of galaxies, as well
as moderate volumes where the internal structures of individual galaxy clusters can be
resolved with very high resolution to investigate, how the presence of baryons and their
associated physical processes like cooling and star-formation are affecting the
systematic difference between mass averaged velocities of dark matter and the ICM 
inside a cluster. We, for the first time, quantify the peculiar motion of galaxy clusters
as function of the large scale environment. We also demonstrate that especially in very 
massive systems, the relative velocity of the ICM compared to the cluster peculiar velocity
add significant scatter onto the inferred peculiar velocity, especially when measurements
are limited to the central regions of the cluster. Depending on the aperture used, this
scatter varies between 50\% and 20\%, when going from the core (e.g. ten percent of the 
virial radius) to the full cluster (e.g. the virial radius). 
\end{abstract}


\begin{figure*}
\includegraphics[width=1.0\textwidth]{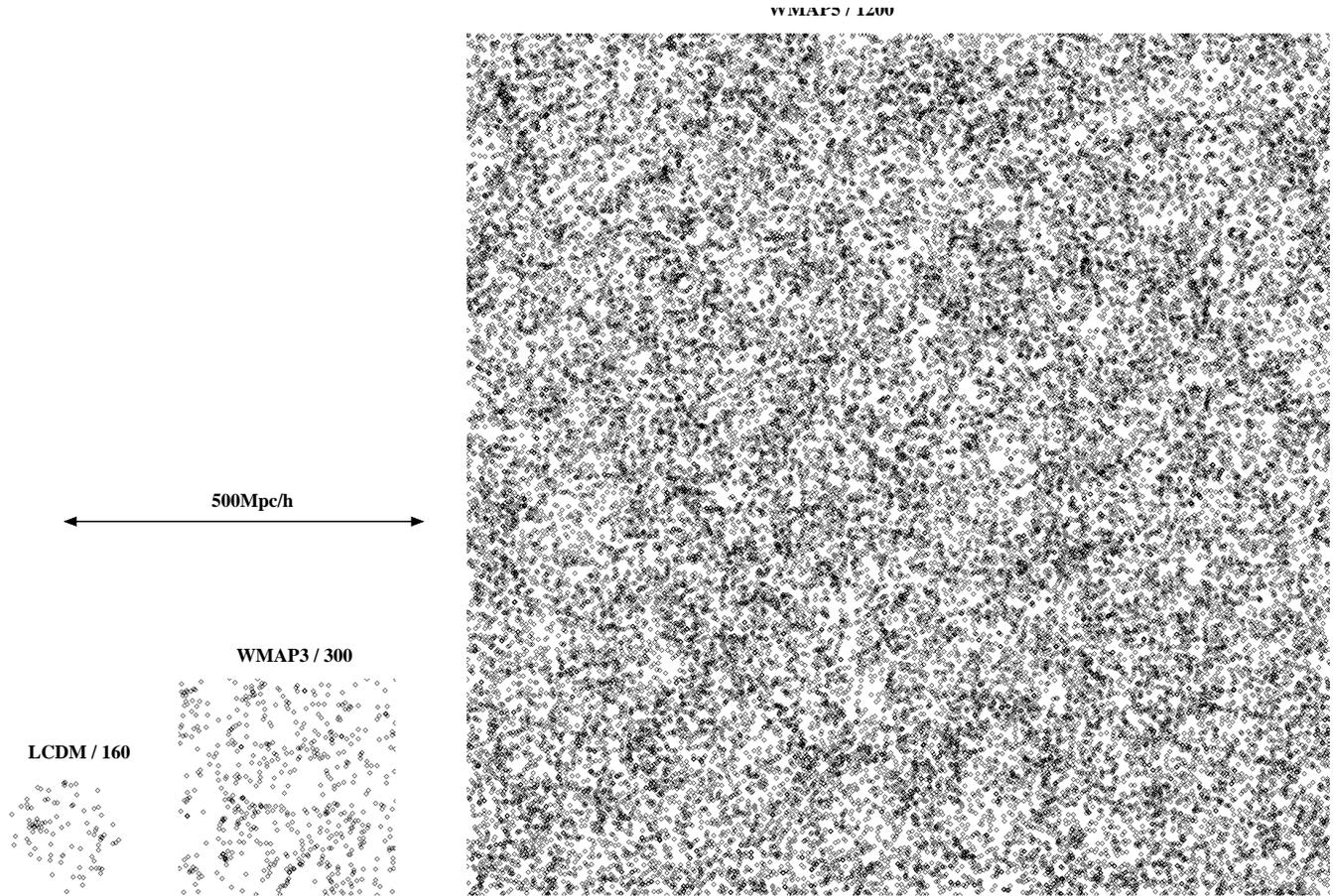}
\caption{Visualization of the 3 different simulations used. Shown are the position of 
all galaxy clusters with masses above $10^{14} M_\odot/h$.}
\label{fig:sim}
\end{figure*}

%##################################################################################
%########################## Introduction ##########################################
%##################################################################################

\section{Introduction} \label{sec:intro}

The increasing sensitivity of the South Pole Telescope, ACT, PLANCK, 
MUSTANG, CARMA, AMIBA, AMI and several other CMB experiments provide the hope that
peculiar velocities of individual clusters of galaxies will be measured
rather soon using the kinematic effect \citep{1980MNRAS.190..413S}. Also next 
generation of X-ray telescopes with microcalorimeters, like the soon coming
Astro-H mission or the planned Athena or SMART-X missions will be able to measure the
kinematic of the intra cluster medium (ICM) with so far unaccomplished precision. 
In fact, recently the Atacama Telescope Team announced their first detection
of the kinematic SZ (kSZ) effect due to the relative motion of the groups of galaxies 
\citep{2012PhRvL.109d1101H}. Thereby, measuring the kinematic SZ effect opens a unique 
and direct method of measuring the peculiar velocities of the the hot ICM in clusters
and groups of galaxies relative to CMB monopole and if associated to the peculiar velocity
of clusters in general probe for the first time an so far unexplored prediction
of our cosmological model (for example recently \citet{2012arXiv1211.0668K} considered
the measurement of the such signal to constrain models of modified gravity).
More precisely, kSZ measurements are able to provide the value of the 
radial component of the ICM peculiar velocity $v_\mathrm{pec,ICM}^r$ relative to 
the local CMB monopole at the position of the cluster of galaxies, 
\begin{equation}
\frac{\Delta T_\mathrm{CMB}}{T_\mathrm{CMB}} = -\frac{\sigma_t}{c} \int n_e v_\mathrm{pec,ICM}^r \mathrm{d}r, 
\end{equation}
folded with the local electron density $n_e$ which needs to be additionally inferred from X-ray 
observations; here $\sigma_T$ and $c$ denote the Thomson cross-section and the speed of light,
respectively. Microcalorimeters on-board of future X-ray satellites will be able to provide 
the redshift of the gas radiating in the iron 6.7 and 6.9 keV lines. This redshift represents
the sum of the Hubble recession velocity $v_\mathrm{Hubble}(z)$ and the radial component of peculiar velocity
of the ICM within the cluster,
\begin{equation}
  v_\mathrm{X-ray} = v_\mathrm{Hubble}(z) + v_\mathrm{pec,ICM}^r 
\end{equation}
At the same time, the broadening of the 6.7 keV line complex will give information about turbulence 
and bulk motions in ICM \citep[see discussion in][]{2003AstL...29..791I,2005MNRAS.364..753D}. 
Therefore combining X-ray and 
kSZ measurements permit to separate the Hubble flow velocity and peculiar velocity of ICM. However,
in galaxy clusters the situation is more complex, as the average radial peculiar velocities of ICM 
and the halo as a whole (including dark matter, ICM and mass in galaxies and field stars) can differ 
significantly. As a result an unavoidable systematic error will appear when the peculiar velocity of
clusters of galaxies as a whole is inferred from observations.

It is well known that both ICM and dark matter in clusters of galaxies
have large bulk motions due to merging of clusters of galaxies
with other clusters or groups of galaxies, infall of fresh dark matter and
ICM in the filaments \citep{1999ApJ...525..554F,2003AstL...29..783S} and 
activity of a black hole in the center of the dominant galaxies 
\citep{2001ApJ...554..261C}. These motions often have velocities of order the sound speed
of the ICM and thereby exceeding several times the expected peculiar velocity 
of the cluster as a whole. Observationally, such bulk motions can be seen
by induced cold fronts or even shocks within the ICM 
\citep[see][for a review]{2007PhR...443....1M}. Such cold fronts suggest that
large regions of clusters (e.g. the whole core) can have systematic velocity
offsets compared to the peculiar velocity of the galaxy cluster and therefore
could significantly affect the kSZ measurements \citep{2003ApJ...597L...1D}.
These velocity offsets are also reflected in the observed offset between
X-ray and weak lensing centers \citep{2012arXiv1204.2743P} as well as the 
displacement of the central galaxy from the mass center of the galaxy 
cluster \citep{2012arXiv1208.1766Z}. From X-ray observations, current measurements 
of the line broadening so far give only upper limits on the turbulent motions within
the ICM \citep{2011MNRAS.410.1797S,2012ApJ...747...32B}, however interpretation of the pressure fluctuations 
observed in the Coma galaxy clusters indicate turbulent velocities at a level of 10\% of the 
sound speed \citep{2004A&A...426..387S}. It has been recently confirmed that the observed
fluctuations in the X-ray surface brightness of Coma would be compatible with
turbulent velocities of several hundreds of km/s \citep{2012MNRAS.421.1123C}. 
Additionally, \citet{2012arXiv1208.2990S} recently presented the evidence for the presence 
of bulk motions far beyond the core regions in the Perseus cluster based on measured 
X-ray surface brightness features.

\begin{figure*}
\includegraphics[width=0.49\textwidth]{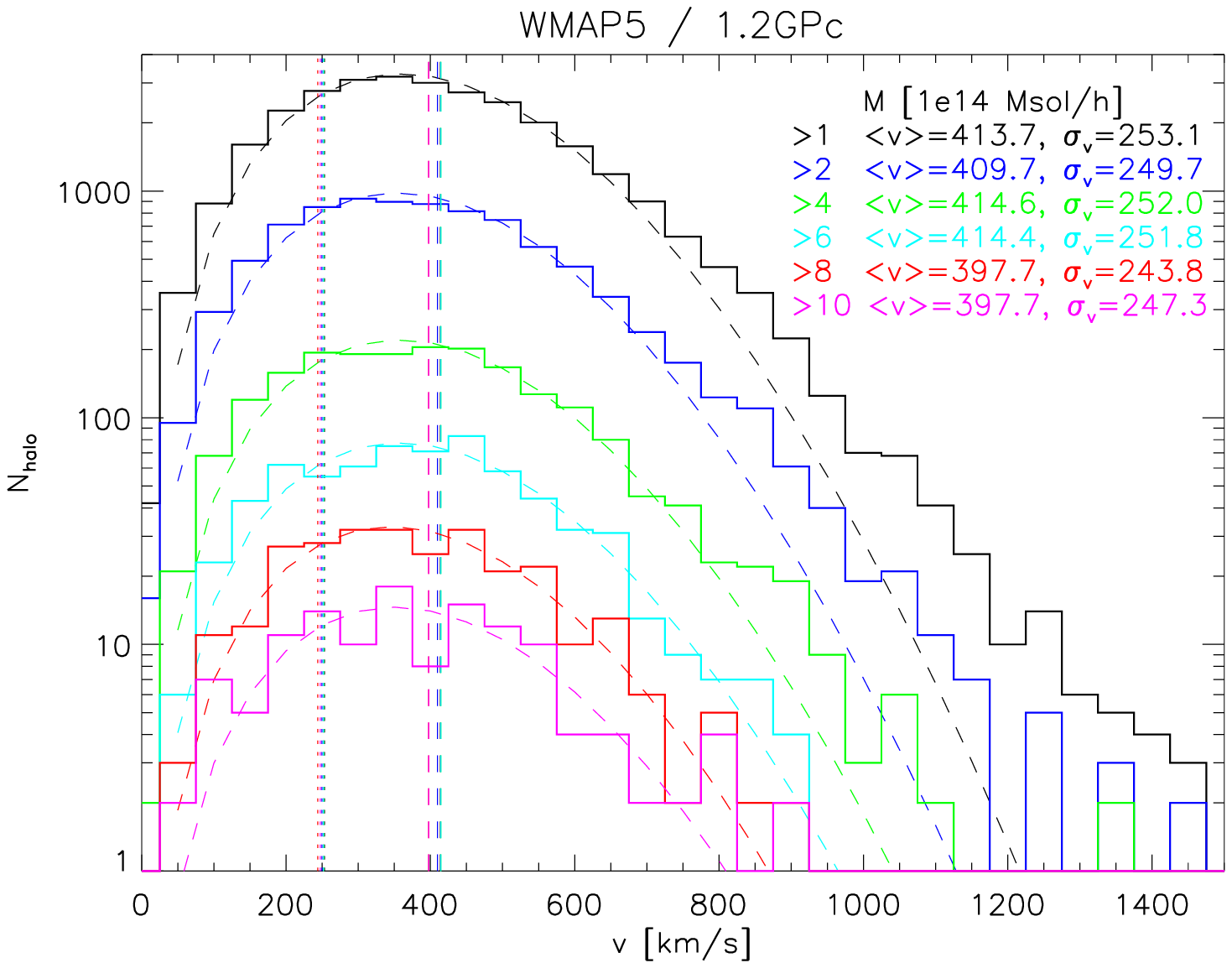}
\includegraphics[width=0.49\textwidth]{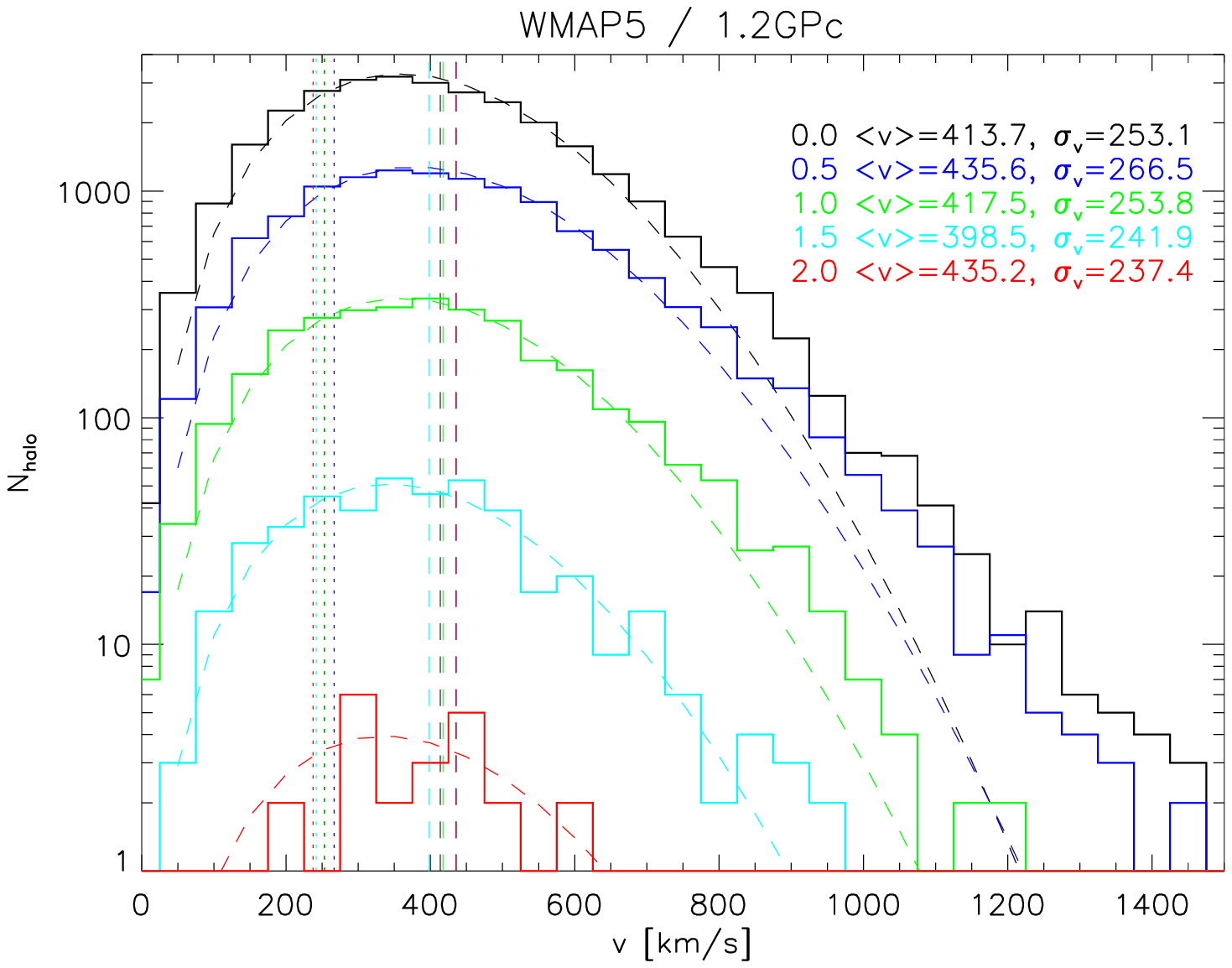}
\caption{Histogram of the  peculiar motion of galaxy clusters from the large cosmological
box {\it WMAP5 / 1200}. Shown in the left panel are the histograms for clusters above $10^{14}$, $2\times10^{14}$,
$4\times10^{14}$, $6\times10^{14}$, $8\times10^{14}$ and $10^{15}$ $M_\odot/h$ whereas the right panel shows
the histograms for clusters above $10^{14}$ $M_\odot/h$ for at redshift $z=2$, $z=1.5$, $z=1$, $z=0.5$ and $z=0$. 
Dashed vertical line marks the median velocity for which values are also quoted in the legends of the panels. 
The dashed line following the histograms is the best fit Maxwell distribution, whereas the vertical dotted 
line marks the standard deviation $\sigma_v$, for which values are also quoted in the legends of the panels.}
\label{fig:halo_vel_hist}
\end{figure*}

Such large bulk motions motions are expected from cosmological simulations 
\citep{2003AstL...29..783S,2003AstL...29..791I} and will drive 
turbulence within the ICM \citep{bryan1998,rasia2004,2005MNRAS.364..753D,
2006MNRAS.369L..14V,2009A&A...504...33V,2011A&A...529A..17V,
2008MNRAS.388.1089I,2011ApJ...726...17P}. Additionally, it has been also shown 
that gravitational perturbations alone (as expected within the cosmological environment of 
clusters) can by them self already induce significant motions within the core of 
clusters \citep[e.g.][]{2006ApJ...650..102A, 2011MNRAS.413.2057R,2010ApJ...717..908Z}. 
Therefore kinematic effect is expected to strongly vary between different regions of the 
cluster. However, using non radiative, cosmological simulations, \citet{2003ApJ...587..524N} 
concluded that the effect of these ICM motions in galaxy clusters is rather mild, 
adding only 50-100 km/s scatter to the observed kSZ signal. Using clusters extracted from
a cosmological simulation which includes the effect of cooling and star-formation, 
\citet{2005MNRAS.356.1477D} find a 10\% to 20\% effect of the internal motion on the
measured peculiar velocity. Additional, they found that the bias between the mass
weighted temperature within the cluster and the X-ray measured temperature can increase 
this error significantly. 

Using numerical simulations of very large cosmological volume, hosting a large number
of rich clusters of galaxies we will study the systematic difference between mass averaged 
velocities of dark matter and the ICM inside galaxy clusters. Therefor we will use both, dark matter
only simulations as well as simulations which are including baryons and their associated physical
processes (like cooling and star-formation) to study their influence on the internal dynamics of 
galaxy clusters. We will thereby demonstrate in detail that this relative velocity between the 
observational signal of the hot ICM and the dark matter within massive clusters of galaxies is 
significantly larger than inferred previously and can introduce an approximately 30\% systematic 
error to the peculiar velocity determination of massive systems.

We will discuss several independent ways to measure this relative velocity using X-ray 
instruments with micro-calorimeters as well as observations of the imprint of the ICM
onto the CMB at radio wavelength, with both should soon become powerful instruments to 
measurements of the bulk motions and turbulence inside clusters of galaxies.

The paper is structured as follows: Section 2 describes the hydrodynamic
simulations used in this paper. Section 3 and 4 generally summarize the peculiar
velocity of galaxy clusters, the influence of baryons and how it is traced by
by different constituents of the cluster. Section 5 the discusses large scale bulk
motions which are present within clusters and how they differ in respect to the 
peculiar velocity of the whole cluster. In Section 6 we demonstrate the 
relative importance of these bulk motions to the observational available signal,
where especially in sub-chapter 6.2 we discuss the different distributions of 
the bulk velocities of dark matter, intra cluster gas and light, and galaxies
within spherical sub-shells at different cluster centric distances as present 
in massive clusters.
In section 7 we show the application to the pairwise velocity signal before we 
summarize our findings in section 8.

%##################################################################################
%########################## Simulations ###########################################
%##################################################################################

\section{Simulations} \label{sec:sim}

The results presented in this paper have been obtained using three different simulations
capturing a wide range in mass and resolution. The largest simulations {\it WMAP5 / 1200}
captures a large enough volume to allow highly significant statistic of peculiar motions
of clusters and super clusters and also allows to explore massive clusters at redshift 
ranges above z=1.
In the medium simulation {\it WMAP3 / 300} , clusters are resolved enough to allow to study 
the systematic differences in the  peculiar motion of the whole clusters and bulk motions of
different parts of the embedded ICM.
In the highest resolution simulation {\it LCDM / 160} gas physics and star-formation
is resolved enough to allow a de-composition of the stellar component and to distinguish
the motion of dark matter in respect to the stars within the central cD, the stars within 
individual galaxies as well as the stars within the so called intra cluster light component.
Figure \ref{fig:sim} gives a visual impression of the different simulations. Shown are the
positions of all clusters above a mass of $10^{14} M_\odot/h$ within the individual simulations. 
If not mentioned otherwise, the simulations where performed with the {\sc{GADGET-3}} code
\citep{2001ApJ...549..681S,2005MNRAS.364.1105S}, which makes use of the entropy-conserving
formulation of SPH \citep{2002MNRAS.333..649S}. These hydrodynamical simulations include 
radiative cooling, heating by a uniform redshift--dependent UV background
\citep{1996ApJ...461...20H}, and a treatment of star formation and feedback processes. 
The latter is based on a sub-resolution model for the multiphase structure of the 
interstellar medium \citep{2003MNRAS.339..289S} with parameters which have been fixed to get 
a wind velocity of $\approx 350$ km/s.

\subsection{LCDM / 160}
The highest resolution simulation used reuses the final output
of a cosmological hydrodynamical simulation of the local universe. Our initial
conditions are similar to those adopted by \citet{Mathis:2002} in their study
(based on a pure N-body simulation) of structure formation in the local
universe.  The galaxy distribution in the IRAS 1.2-Jy galaxy survey is first
Gaussian smoothed on a scale of 7 Mpc and then linearly evolved back in time
up to $z=50$ following the method proposed by \cite{Kolatt:1996}. The
resulting field is then used as a Gaussian constraint \citep{Hoffman1991} for
an otherwise random realization of a flat $\Lambda$CDM model, for which we
assume a present matter density parameter $\Omega_\mathrm{0m}=0.3$, a Hubble constant
$H_0=70$ km/s/Mpc and a r.m.s. density fluctuation $\sigma_8=0.9$.  The volume
that is constrained by the observational data covers a sphere of radius $\sim
80$ Mpc/h, centered on the Milky Way. This region is sampled with more than 50
million high-resolution dark matter particles and is embedded in a periodic
box of $\sim 240$ Mpc/h on a side. The region outside the constrained volume 
is filled with nearly 7 million low-resolution dark matter particles, allowing 
a good coverage of long-range gravitational tidal forces.

Unlike in the original simulation made by \citet{Mathis:2002}, where only the
dark matter component is present, here we followed also the gas and
stellar component. For this reason we extended the initial conditions by
splitting the original high-resolution dark matter particles into gas and dark
matter particles having masses of $m_\mathrm{gas} \approx 0.48 \times 10^9\; M_\odot/h$
and $m_\mathrm{dm} \approx3.1 \times 10^9\; M_\odot/h$, respectively; 
this corresponds to a cosmological baryon
fraction of 13 per cent. The total number of particles within the simulation
is then slightly more than 108 million and the most massive clusters is
resolved by almost one million particles.
The gravitational force resolution (i.e. the comoving softening length) of
the simulation has been fixed to be 7 kpc/h (Plummer-equivalent), fixed in
physical units from z=0 to z=5 and then kept fixed in comoving units at higher
redshift.

In addition, this simulation follows the pattern of
metal production from the past history of cosmic star formation
\citep{2004MNRAS.349L..19T,2007MNRAS.382.1050T}. This is done by computing the contributions 
from both Type-II and Type-Ia supernovae and energy feedback and
metals are released gradually in time, accordingly to the appropriate
lifetimes of the different stellar populations. This treatment also
includes, in a self-consistent way, the dependence of the gas cooling on
the local metalicity. The feedback scheme assumes a Salpeter IMF
\citep{1955ApJ...121..161S}, and its parameters have been fixed to get 
a wind velocity of $\approx 480$ km/s.  

More detailed information regarding this simulation and the technique to
distinguish between the satellite galaxies, the cD and the ICL component 
can be found in Dolag et al. 2010.

\subsection{WMAP3 / 300}

As medium size cosmological simulation we used a cosmological box of size 
$(300 {\rm{Mpc}}/h)^{3}$, resolved with $(768)^{3}$ dark matter particles 
with a mass of $m_\mathrm{dm} \approx 3.7 \times 10^{9} {\rm{M_{\odot}}} h^{-1}$ and 
the same amount of gas particles, having a mass of 
$m_\mathrm{gas} \approx 7.3 \times 10^{8} {\rm{M_{\odot}}} \ h^{-1}$. 
Here we use the concordance $\Lambda$CDM model, adapted to the WMAP3 
values \citep{2007ApJS..170..377S}, which assumes a matter density of 
$\Omega_\mathrm{0m}=0.268$, a baryon density of $\Omega_\mathrm{0b}=0.044$, a Hubble 
parameter $h=0.704$, a power spectrum normalization of $\sigma_{8}=0.776$
and a spectral index of $n_{s}=0.947$. 
More information on this simulation can be found in 
\citet{2011MNRAS.415.2758D,2012arXiv1205.3163D} and \citet{2012MNRAS.419.1588F}.

\begin{figure}
\includegraphics[width=0.5\textwidth]{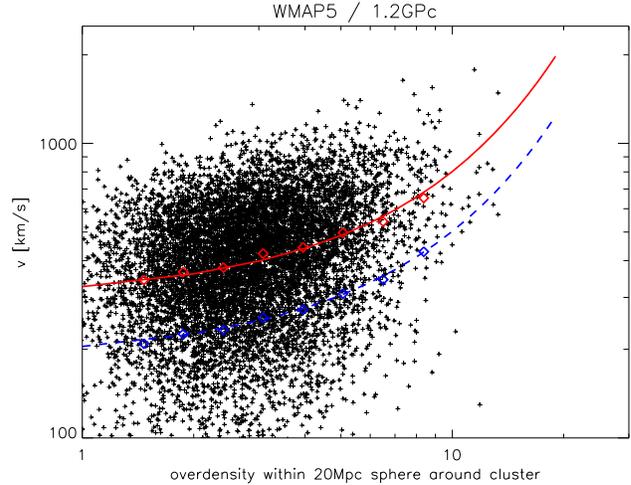}
\caption{Dependence of the cluster peculiar motion on the large scale environment. Shown are the
 peculiar motions of all clusters above $2\times10^{14}$ $M_\odot/h$ at z=0 from the large cosmological box 
{\it WMAP5 / 1200} as function of the local over-density within 20 Mpc around each cluster. 
Clear to see that the median peculiar velocity (red diamonds, solid line) and similarly the
best fit standard deviation $\sigma_v$ (blue diamonds, dashed line) rises by a factor of two between 
low density and high density environments.} 
\label{fig:halo_vel_environment}
\end{figure}

\subsection{WMAP5 / 1200}

As large cosmological simulations we considered a box of (1200 Mpc/$h)^3$, 
resolved with $960^{3}$ dark matter particles with a mass of
$m_{\rm{dm}} \approx 1.2 \times 10^{11} h^{-1} \rm{M}_\odot$ and the same
amount of gas particles, having a mass of $m_{\rm{gas}} \approx 2.4 \times 10^{10} h^{-1} \rm{M}_\odot$. 
The gravitational force has a Plummer-equivalent softening length of $\epsilon_{l}=25 h^{-1}$ kpc. 
This simulations assumes a `concordance' $\Lambda$CDM model, where we fix the relevant parameters
consistently with those derived from the analysis of the WMAP 5-year data \citep{2009ApJS..180..330K}: 
$\Omega_\mathrm{m0}=0.26$ for the matter density parameter,
$\Omega_{\Lambda0}=0.74$ for the $\Lambda$ contribution to the density
parameter, $h=0.72$ for the Hubble parameter (in units of $100$ km
s$^{-1}$ Mpc$^{-1}$). The initial power spectrum has a spectral index 
$n=0.96$ and is normalized in such a way that $\sigma_{8}=0.8$. 
More information on this simulation can be found in
\citet{2009MNRAS.398..321G,2010JCAP...07..013R,2010MNRAS.402..923R},
where the dark matter only counterpart of these simulations was used. 

\begin{figure*}
\includegraphics[width=1.0\textwidth]{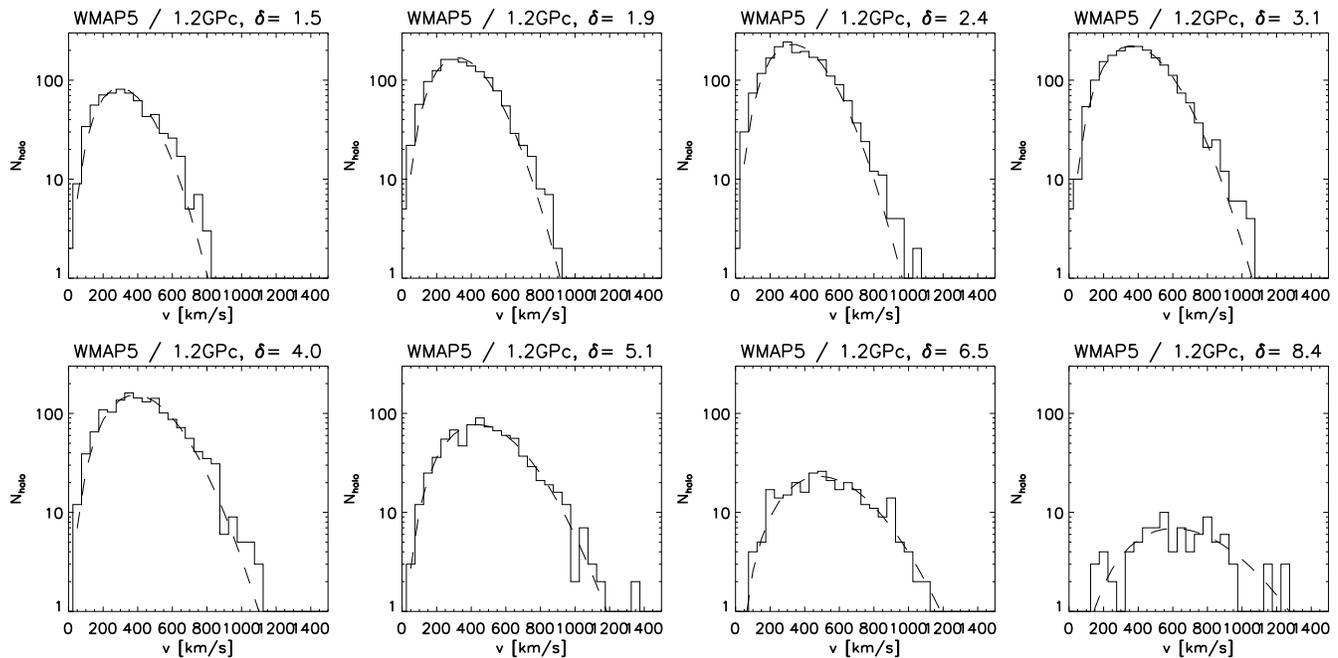}
\caption{Distribution of the peculiar velocities of the clusters from figure \ref{fig:halo_vel_environment}
as function of the environmental over-density (solid lines) and the best fit to a Maxwell distribution 
(dashed lines). }
\label{fig:halo_vel_hist_environment}
\end{figure*}

\subsection{Post processing}

We are using {\sc SUBFIND} \citep{2001MNRAS.328..726S, 2009MNRAS.399..497D, 2010MNRAS.405.1544D}
to define halo and sub-halo properties. {\sc SUBFIND} thereby identifies substructures 
as locally overdense, gravitationally bound groups of particles. Starting with a 
halo identified through the Friends-of-Friends algorithm, a local density is estimated
for each particle with adaptive kernel estimation using a prescribed
number of smoothing neighbours. Starting from isolated
density peaks, additional particles are added in sequence of decreasing
density. Whenever a saddle point in the global density field
is reached that connects two disjoint overdense regions, the smaller
structure is treated as a substructure candidate, followed by merging
the two regions. All substructure candidates are subjected to an
iterative unbinding procedure with a tree-based calculation of the
potential. The {\sc SUBFIND} algorithm is discussed in full in \citep{2001MNRAS.328..726S}
and its extension to dissipative hydrodynamical simulations
that include star formation in \citet{2009MNRAS.399..497D}.
Based on a differences in the dynamics, the unbinding procedure for the central
galaxy of each halo can be modified, which leads to a spatial separation of the 
two, dynamically well separated components into a cD and a diffuse stellar 
component (DSC). For details see \citet{2010MNRAS.405.1544D}.

The Mass and Radius of our clusters will refer to a spherical over-density
according to a density contrast of 200 with respect to the mean density,
which we associate with the viral properties. Synthetic maps
of observables are produced using {\sc SMAC} \citep{dolag2005}. If not stated differently,
we will call peculiar velocity the mean (and mass weighted among all constituents) 
velocity of a halo, averaged within the virial radius while we will generally refer
to bulk motions if we speak about motions of certain components averaged over
sub-regions within the cluster.

%##################################################################################
%########################## Cluster Bulk ##########################################
%##################################################################################

\section{Cluster peculiar motions} \label{sec:cluster_bulk}

The distribution of the overall peculiar motions of galaxy clusters reflect the large scale
velocity field, which still probes the linear regime of structure formation. Therefore
they are expected to follow a Maxwell distribution \citep{1994ApJ...436...23B}.
In this section we discuss the cluster peculiar motion as extracted from the large, cosmological
simulation ({\it WMAP5 / 1200}).

\subsection{Mass and redshift dependence} \label{sec:cluster_bulk1}

For the current (low mass density) cosmology the typical (e.g. median) cluster peculiar motion is 
expected to of order of 400 km/s \citep{1994ApJ...436...23B}. Figure \ref{fig:halo_vel_hist},
shows the distribution of cluster peculiar velocities for different cut in cluster mass as well 
as for different redshifts. Independently on the actual cluster mass, or the redshift, the 
median peculiar velocities is $\approx 410$ km/s. They mainly follow an Maxwell distribution, 
\begin{equation}
P(v_\mathrm{pec}) = A_0 v_\mathrm{pec}^2 \mathrm{exp}\left(\frac{-v_\mathrm{pec}^2}{2\sigma_v^2}\right) 
\end{equation}
as expected \citep{1994ApJ...436...23B}, but with an excess at higher velocities (see next 
sub-section for a discussion of the origin of that excess). For a perfect, Maxwell distribution, 
the mean velocity $\left<v_\mathrm{pec}\right>$ is related to the standard deviation $\sigma_v$ by 
$\left<v_\mathrm{pec}\right> = \sqrt{8/\pi} \sigma_v$, which in this case gives a typical standard 
deviation of $\sigma_v \approx 250$ km/s. 

\subsection{Environment dependence of peculiar motions} \label{sec:cluster_bulk_environment}

To explain the origin of the excess of larger peculiar velocities we investigated the
 peculiar motion of the clusters as extracted from the large, cosmological simulation 
({\it WMAP5 / 1200}) as function of the local over-density. Therefore, we selected
all clusters above a mass limit of $2\times10^{14}$ $M_\odot/h$ at z=0 and calculated the
over-density of matter within a 20 Mpc sphere around each cluster. 
Figure \ref{fig:halo_vel_environment} shows the distribution of the peculiar velocities
as function of this large-scale over-density. Clear to see is the trend to have larger
peculiar motions of clusters in higher density region, which boost the median cluster
peculiar velocity by almost a factor of two in the high density regions. We also calculated 
the velocity distribution by binning the clusters by the over-density of the environment.
In this case, the individual distributions follow very closely a Maxwell distribution without
any excess at large velocities, as shown in figure \ref{fig:halo_vel_hist_environment}, where
the velocity distribution and the best fit Maxwell distribution are shown for the individual
over-density bins. The best fit value for the standard deviation $\sigma_v$ (blue diamonds 
in figure \ref{fig:halo_vel_environment}) follows a functional form of
\begin{equation}
   \sigma_v(\delta_{20\mathrm{Mpc}}) = 185 \mathrm{km/s} \times \mathrm{exp}\left(\delta_{20\mathrm{Mpc}}/10\right)
\end{equation}
which is shown as dashed line in figure \ref{fig:halo_vel_environment}. The mean peculiar velocity follow
(as expected) exactly the same trend but multiplied by $\sqrt{8/\pi}$ as shown by the red diamonds and the
solid line in figure \ref{fig:halo_vel_environment}.

\begin{figure}
\includegraphics[width=0.5\textwidth]{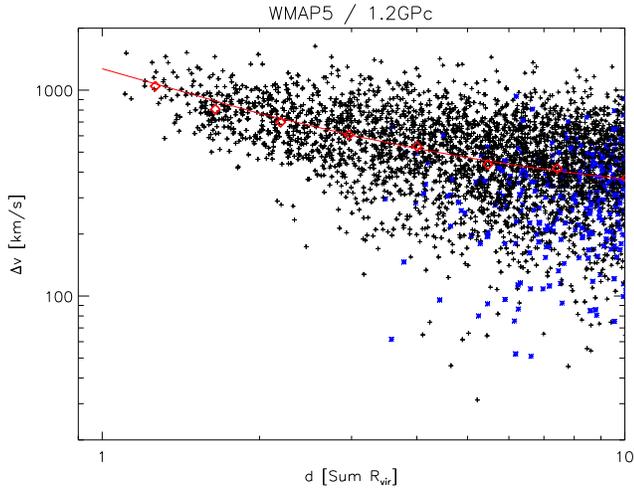}
\caption{Dependence of the relative motions of clusters pairs on their distance. Shown are the
all clusters above $2\times10^{14}$ $M_\odot/h$ at z=0 from the large cosmological box {\it WMAP5 / 1200}.
Crosses (black) are approaching systems, stars (blue) are systems which move away from each other. 
Clear to see that cluster pairs have an strong excess in their pairwise velocity compared
to typical peculiar motions with even stronger signal for close pairs. Diamonds (red) are the median,
pairwise velocity and the best power-law fit (line).}
\label{fig:pairwhise_velocity}
\end{figure}

The velocity difference of close pairs of galaxy clusters can also be used to boost the measurement
of cluster peculiar motions \citep{2000ApJ...533L..71D}. Figure \ref{fig:pairwhise_velocity} shows how the velocity difference
between cluster pairs is boosted due to the merging process of large scale structure. Here, the distance $d$ is
measured in units of the sum of the virial radii of the two clusters, meaning that $d=1$ on the x-axis
means that the virial radii of the two clusters touch each other. It is worth to mention that only at larger
distances (e.g. $d > 3$) we find systems which actually move away from each other (blue stars) in figure 
\ref{fig:pairwhise_velocity}, whereas close by systems always approach each other. This means that (as expected),
violent relaxation for clusters is effective enough to do not allow a merging system to completely detach
again (e.g. $d>1$). Of course, within the virial radius the merging process can produce gas rich sub-structures
which might move in opposite directions, as observed in systems like the {\it bullet cluster} \citep{2004ApJ...606..819M}. 
The median velocity here can easily exceed 1000 km/s (red diamonds in figure \ref{fig:pairwhise_velocity}) and therefore 
is 2.5 times larger than the median peculiar velocity of individual clusters. The solid line
marks the simple functional form of
\begin{equation}
\left<v(d)\right> = 270 \mathrm{km/s} + 1000 \mathrm{km/s} \times d^{-1}
\end{equation}   
shown a solid (red) line in figure \ref{fig:pairwhise_velocity}.

%##################################################################################
%########################## ICM Bulk ##############################################
%##################################################################################

\section{The influence of baryons} \label{sec:icm_bulk}

In this section we want to discuss the effect of baryonic
material on the overall peculiar motions of galaxy clusters
and especially the peculiar motions traced by the different
components within galaxy clusters like dark matter, ICM and stars.
 As already mentioned, the distribution of the overall peculiar motions of 
galaxy clusters reflect the large scale velocity field, which still probes 
the linear regime of structure formation and therefore we do not
expect the peculiar velocities to be effected by the presence of
the baryonic component. This is demonstrated in detail in the Appendix A1, 
figure \ref{fig:halo_vel_hist_mass_gas}.

\begin{figure}
\includegraphics[width=0.5\textwidth]{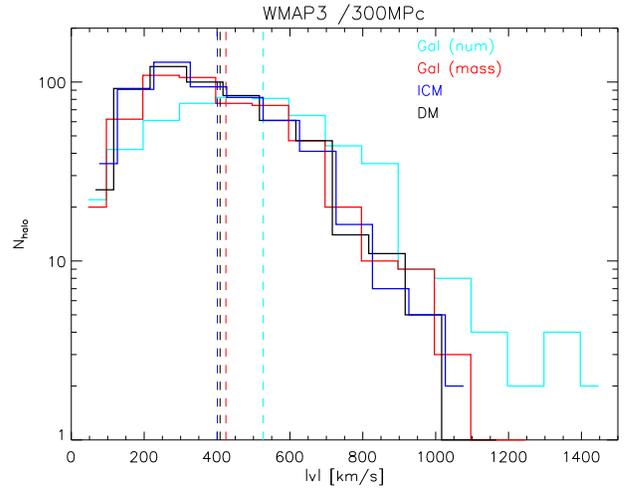}
\caption{Histogram of the peculiar motion of galaxy clusters from the medium size cosmological
box {\it WMAP3 / 300} at $z=0$. Shown are the histograms for clusters above $10^{14}$ $M_\odot/h$ 
for the dark matter (black line), ICM (blue line) and galaxy component (red line for mass weighted, 
light blue line for number weighed).}
\label{fig:halo_vel_hist_components}
\end{figure}

Although the cluster peculiar velocities (recap that those is the mass weighted mean
over all constituents of the cluster) does not change significantly when baryonic physics
is included it is interesting to see, if the mean velocity of the different individual
constituents within the cluster differs. As we continue to look at mean values within the
virial radius, we will still call them peculiar velocities. As we also want to include 
sub-structure, representing galaxies, we will switch here to the medium size cosmological box
({\it WMAP3 / 300}) as here we have good enough resolution to resolve at least 10 member
galaxies in a $10^{14}$ $M_\odot/h$ halo and several hundred of member galaxies in the most 
massive systems. Figure \ref{fig:halo_vel_hist_components} shows the peculiar velocity 
distribution as traced by the different components. There is not much difference between the 
distribution (and the median values) of the peculiar motion measured by the dark matter and 
the ICM component within the cluster. The ICM velocity match almost the peculiar velocity 
(within 1\%). However, the peculiar velocity traced by the galaxies 
is slightly biased high (ca. 5\%) when using the mass weighted mean and strongly biased 
(ca. 12\%) when one calculate the peculiar velocity as the pure mean of the
velocity of all galaxies. Obviously, when weighting the velocities of the individual galaxies 
by their (stellar) mass gives the central galaxy (which in this case additional inherits a 
large stellar component in form of the diffuse stellar component) a very large weight.

\begin{figure*}
\includegraphics[width=1.0\textwidth]{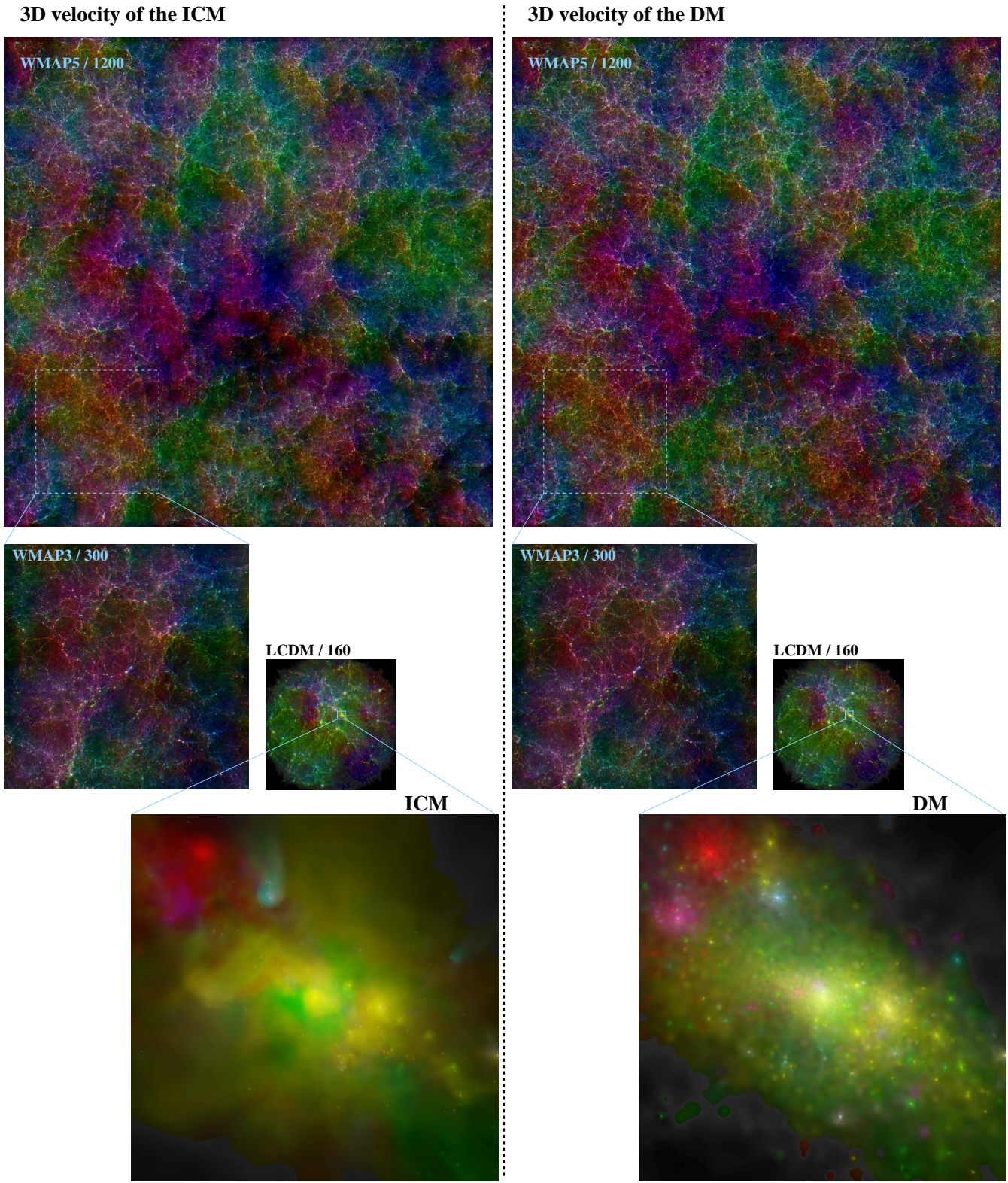}
\caption{Visualization of velocity field the 3 different simulations using the 
the ray--tracing software {\small SPLOTCH}. Left side shows the ICM component,
right side shows the dark matter component. The color composition reflect the
three dimensional velocity field (see text for details), whereas the intensity 
reflects the density.}
\label{fig:sim2}
\end{figure*}

%##################################################################################
%########################## ICM Bulk ##############################################
%##################################################################################

\section{Relative ICM bulk motion} \label{sec:icm_bulk}

\begin{figure*}
\includegraphics[width=0.49\textwidth]{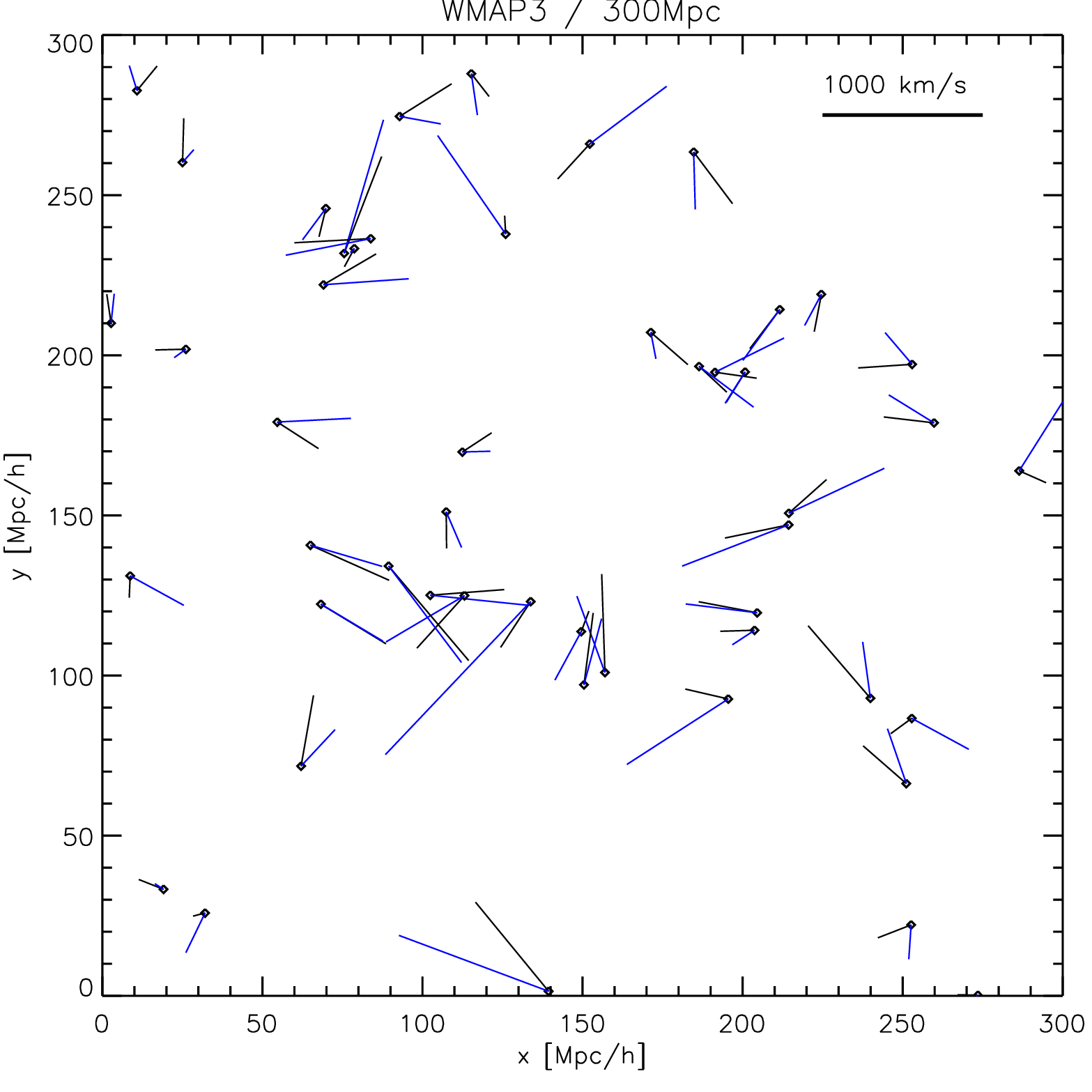}
\includegraphics[width=0.49\textwidth]{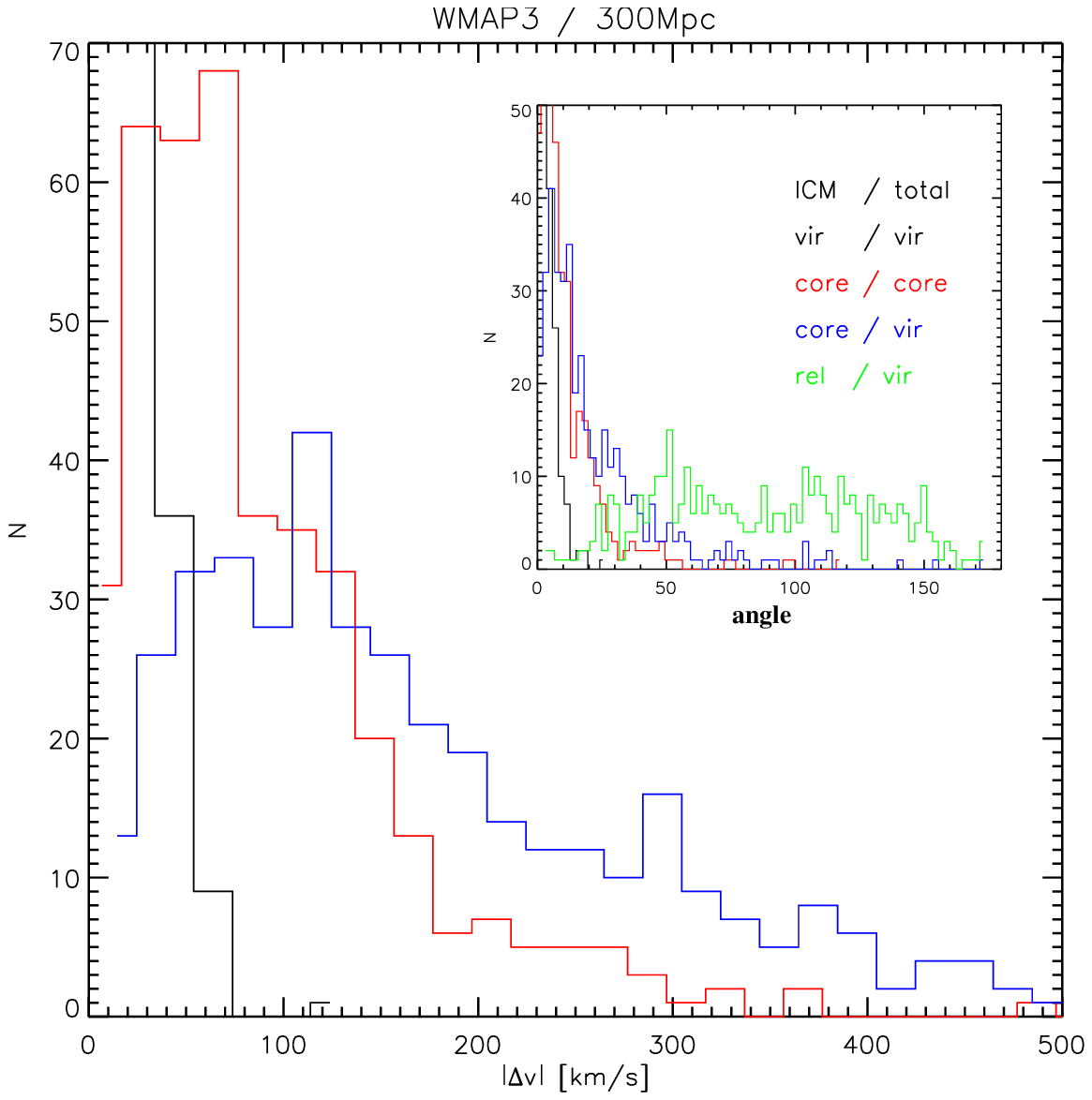}
\caption{Shown are the different velocities of the different components within a halo. 
Left panel shows the velocity vectors of the ICM core compared to the mean velocity 
of the whole halo. Right panel shows the distribution of the angle between the velocity
of the ICM component compared to the total halo measured within different regions.}
\label{fig:halo_vec}
\end{figure*}

While \citet{2003ApJ...597L...1D} speculated from observation of cold fronts in galaxy 
clusters, that the baryonic motion (e.g. sloshing) inside of galaxy clusters could
significantly affect the kSZ measurements, \citet{2003ApJ...587..524N} used non
radiative simulations to conclude that the effect of the ICM motions in galaxy clusters
is rather mild, adding only 50-100 km/s scatter to the observed kSZ signal. Here we want
to re-examine the amount of sloshing using up to date cosmological simulations which 
include the effect of cooling and star-formation extending previous studies by 
\citet{2005MNRAS.356.1477D} to more massive systems present in our simulations.

To give an impression of the cosmological velocity field reflected by the different
components in our simulations we used the {\small SPLOTCH} package \citep{Dolag08}, 
a Ray--tracing visualization tool for SPH simulations. Instead of mapping a scalar
value to a color table, we mapped the three velocity components directly into the 
RGB colors. To obtain a logarithmic like scaling but preserve the sign of the 
individual velocity components we used a mapping based on the inverse of the 
hyperbolic sin function, e.g.  $\left<r,g,b\right> \propto \mathrm{asinh}(v_{x,y,z})$. 
For the DM particles we calculated a smoothed density field based on all DM 
particles using a SPH like kernel function. The resulting Ray--tracing images, 
are shown in Figure~\ref{fig:sim2}. The left column shows the result for the
ICM component, whereas the right column shows the dark matter component.
Clear to see that the large scale velocity field is equally represented in 
the dark matter as well as ICM component, as expected. However, if one zoom
into a galaxy cluster, there are striking differences in the velocity field of the
two components. On one hand, within the dark matter there are more substructures 
(e.g. galaxies) which have individual velocities. On the other hand, and more
importantly, there are quite striking, large scale velocity structures within 
the ICM visible, which are not reflected in the dark matter.

This is expected from the observations of ICM sloshing within the cluster potential,
where velocities of up to 1000 km/s are inferred. Such velocities are quite comparable 
with the sloshing velocities predicted by our simulations, as will be discussed in detail in
section \ref{sec:icm_bulk_profile}.

\subsection{Relative motions in the core} \label{sec:icm_bulk_core}

As a first step, we can divide the cluster into two parts, namely everything within the
viral radius $R_\mathrm{vir}$ and the core (which for simplicity we define simple as the
region within $0.1\times{}R_\mathrm{vir}$) where we will subscribe the quantities calculated
within regions with {\it vir} and {\it core} respectively. For those two regions, we then
can define the mass weighed mean velocity of all constituents or the mass weighed velocity
of the baryonic component, which we will subscribe with {\it total} and {\it ICM} respectively. 
Therefore $v_\mathrm{vir}^\mathrm{total}$ correspond to the peculiar velocity of the halo,
whereas $v_\mathrm{core}^\mathrm{ICM}$ correspond more to the bulk of the signal expected
from measurements.
% of kSZ or using X-ray spectrometers.

Figure \ref{fig:halo_vec} shows the difference of those velocities for the medium size
cosmological box ({\it WMAP3 / 300}). The left panel show the position and velocity vectors
for the 50 most massive clusters. Shown is the peculiar velocity (black lines) compared
to the  bulk of the signal expected from measurements as represented by $v_\mathrm{core}^\mathrm{ICM}$ 
(blue lines). Clear to see that amplitude (e.g. length of the lines, the corresponding amplitude is 
given in the legend in the upper right part of the plot) as well as direction are not 
the same and in the majority of the cases the velocity of the ICM in the center is larger than the
peculiar velocity of the cluster. In some cases the direction can bee even completely opposite.
The right panel shows a more quantitative result by showing the distribution of the length (main plot)
as well as the angle (inlay) between the different velocities for all clusters above a mass 
limit of $10^{14}$ $M_\odot/h$.
The tightest correlation (e.g. largest peak at small angles) is found when comparing
$v_\mathrm{vir}^\mathrm{ICM}$ with $v_\mathrm{vir}^\mathrm{total}$ indicating that the overall
mass weighted velocity of the ICM within the virial radius is typically within 50 km/s and 5 degrees
of the peculiar velocity (black histogram). The difference between the motion of the total
mass within the core and the ICM within the core (e.g. angle between $v_\mathrm{core}^\mathrm{ICM}$ and 
$v_\mathrm{core}^\mathrm{total}$) is found to be typically different by less then 100 km/s and 20 degrees (red line).
The ICM velocity of the core typically is only weakly related to the peculiar velocity, 
typically 100 km/s but with a very pronounced tail to much larger values and the angle between 
$v_\mathrm{core}^\mathrm{ICM}$ and $v_\mathrm{vir}^\mathrm{total}$ 
encompasses typically 30 degree (blue line). The difference between the relative ICM motion inside the core 
and in respect to the peculiar velocity and the peculiar velocity itself is completely
uncorrelated, as can be seen from the flat distribution of the angle (green line). 

We conclude that overall statistically the bulk motion of the ICM (within the virial
radius) is very similar than that one of the dark matter (see figure \ref{fig:halo_vel_hist_components}),
however there is a significant relative motion of the ICM in respect to the dark matter
in nearly every cluster, which is more pronounced within the core radius of the cluster
than in the outskirts. Here we want to remind that the X-ray line emissivities are 
proportional to the square of the electron density and to the abundance of iron.
Therefore the central part of the cluster will contribute to the bulk of the total 
line emission signal. On the other hand, kSZ measurements are proportional to the 
column density of electrons and therefore contains a significant contribution 
from the gas in the periphery of clusters.

\begin{figure}
\includegraphics[width=0.49\textwidth]{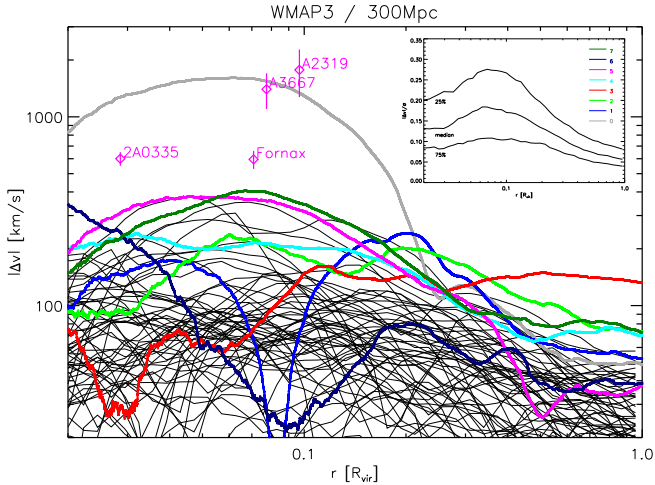}
\caption{Radial profile (cumulative) of the velocity difference between the ICM and the
dm component. Shown are every tenth profiles from all clusters above $10^{14}$ $M_\odot/h$ taken
from the medium size cosmological box {\it WMAP3 / 300} at $z=0$. The profiles of the 8 most 
massive clusters in the simulation are shown as extra, color coded lines. The data points are 
inferred motions of the core as imprinted in cold fronts in observed clusters.
The right inlay shows the median and the 25/75 percentile when scaling the velocities of the 
individual clusters by the virial velocity dispersion calculated from their virial mass. 
For the typical mass fraction (relative to the total mass) of the different components
as function of cluster centric distance see figure \ref{fig:cluster_composition}.}
\label{fig:vel_prof_045}
\end{figure}

\subsection{Radial profiles of relative motions} \label{sec:icm_bulk_profile}

To quantify the effect of the ICM sloshing inside the cluster potential as expected from
cosmological simulations more quantitatively, we compute the profiles of the velocity 
difference between the ICM and the dark matter within a given radius for all clusters
more massive than $10^{14}$ $M_\odot/h$ from the medium size cosmological box {\it WMAP3 / 300}.
The individual profiles are shown in the left panel of figure \ref{fig:vel_prof_045}.
Clear to see that such velocity difference at distance of the core radius is typically of
order of 100 km/s with many systems reaching velocity differences of several hundreds 
of km/s, the most extreme ones even more than 1000 km/s. This sloshing can be compared with 
the velocities inferred from the observations of cold fronts in real clusters. As comparison,
we over-plot the values for 4 known systems taken from \citet{2007PhR...443....1M} and references
therein. These data points 
should (and actually do quite well) mark the extreme velocity differences to be expected in 
the simulated clusters and therefore the sloshing of the ICM in the cores of our simulated 
clusters is found to fairly represent the amount of sloshing found in observed systems.

The amount of sloshing is expected to be related to the depth of the potential well of the 
cluster and indeed, when scaling the velocity with a value of the virial velocity dispersion
\begin{equation}
\sigma_v = \sqrt{\frac{G M_\mathrm{vir}}{3 R_\mathrm{vir}}}
\label{eq:sigma_v}
\end{equation}
expected for the different clusters according to their virial mass, the scatter in the individual
profiles is strongly reduced. The inlay panel of figure \ref{fig:vel_prof_045} shows the median 
and the 25/75 percentile of the scaled profiles. We therefore find that the cores of the
simulated galaxy clusters
typically are sloshing with approximately 18\% of the virial velocity dispersion, indicating 
that for very massive clusters this signal significantly contributes to the measurement 
of the peculiar velocity, typically 30\% for systems with masses above $10^{15}$ $M_\odot/h$. 
This contribution drops to less than 10\% for systems with masses around $10^{14}$ $M_\odot/h$.
Therefore stacking many small systems as done in \citet{2012PhRvL.109d1101H} indeed might be 
a good choice to minimize the bias induced by the relative ICM motions within the clusters 
and groups.

\begin{figure}
\includegraphics[width=0.5\textwidth]{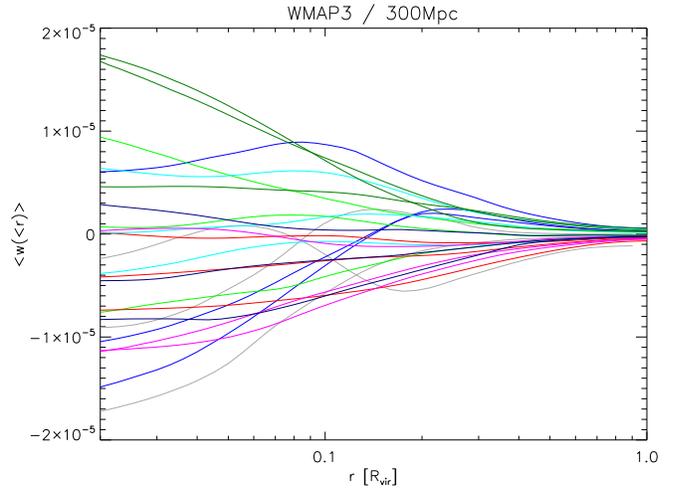}
\caption{Radial profile of kSZ for the 8 most massive halos, (x,y,z) projection
each. Clear to see that the weighting by density gives very large weight to the
contribution by the velocity within the core region, where the difference between 
the local motion of the ICM and the bulk motion of the cluster usually is largest. 
Color of the lines is the same as in figure \ref{fig:vel_prof_045}.}
\label{fig:ksz_profile}
\end{figure}

%##################################################################################
%########################## kSZ ###################################################
%##################################################################################

\begin{figure*}
\includegraphics[width=0.49\textwidth]{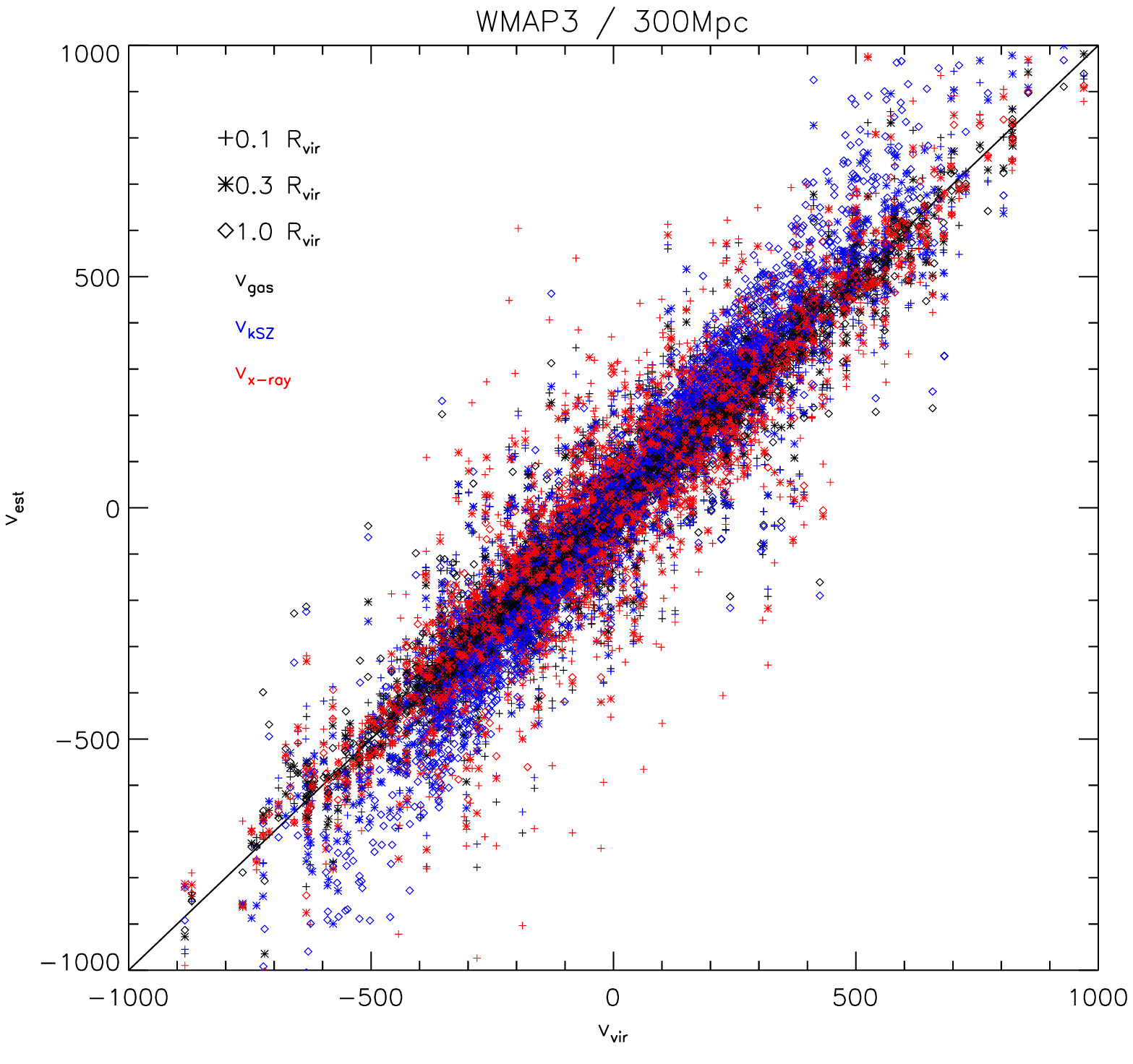}
\includegraphics[width=0.245\textwidth]{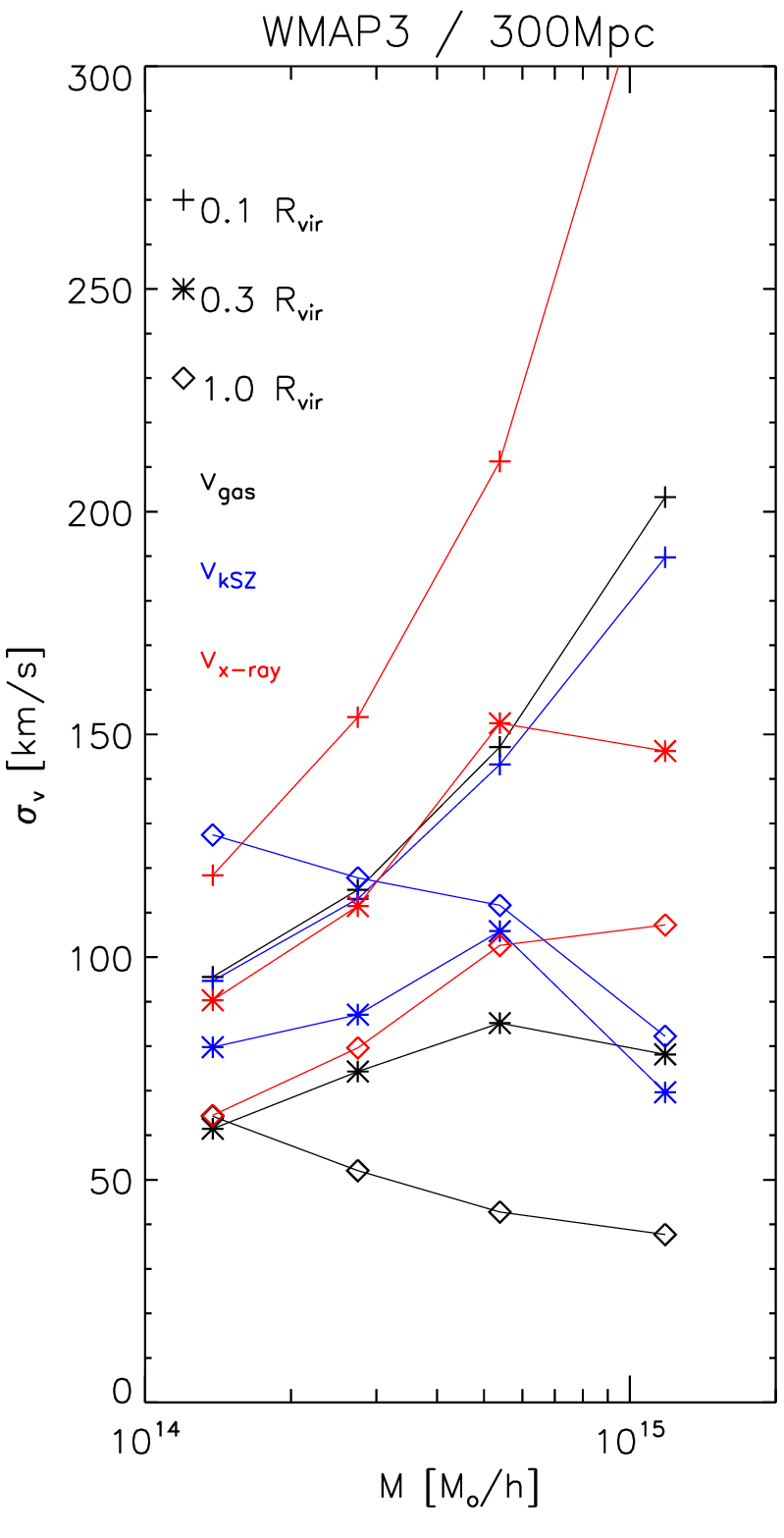}
\includegraphics[width=0.245\textwidth]{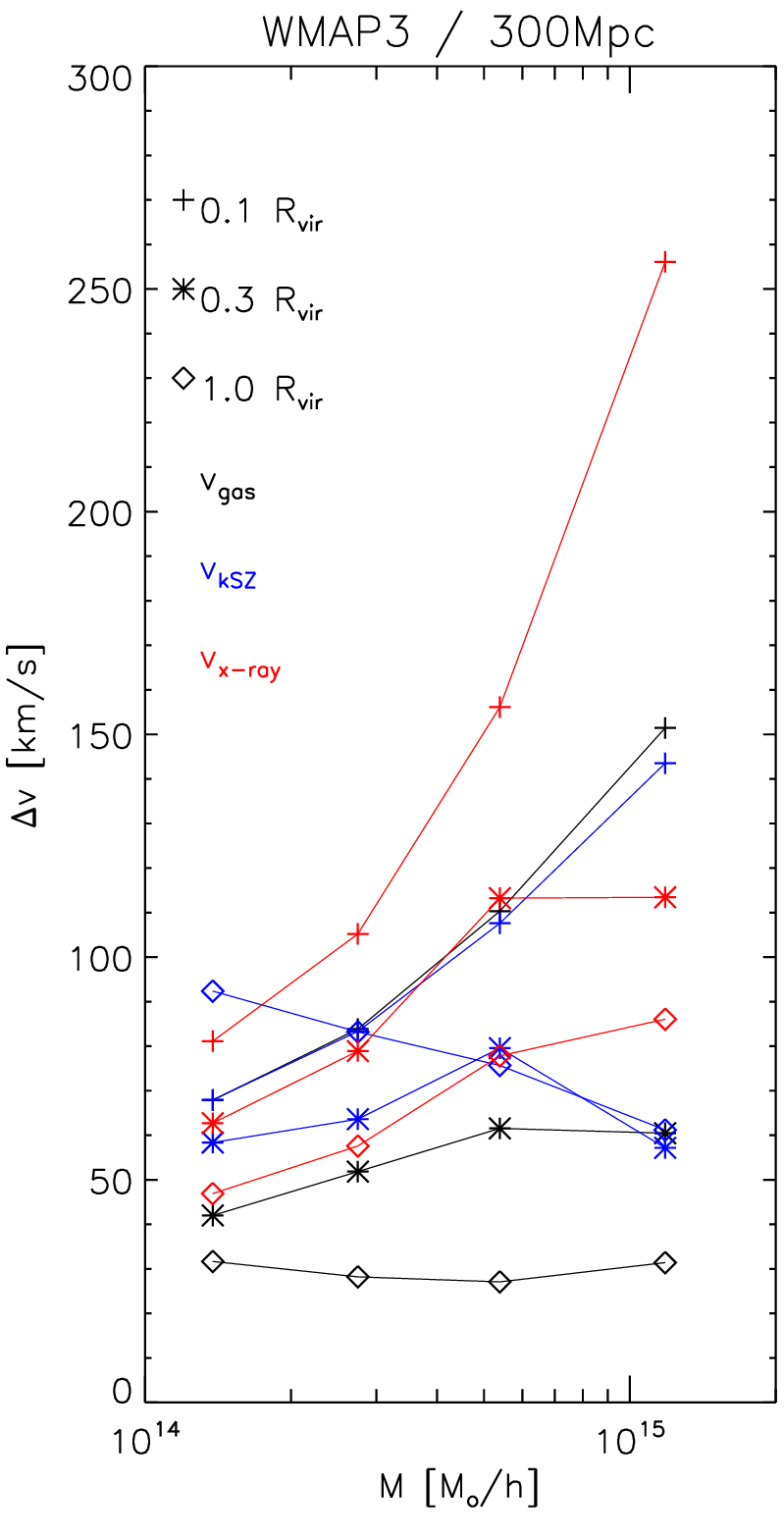}
\caption{Difference between the peculiar velocity along the line of sight
and the velocity inferred from different observables. Left panel shows a 
one-by-one scatter plot of all halos using the x,y and z projection of 
each halo. Different symbols refer to different aperture sizes (crosses = core, 
stars = $0.3\times{}R_\mathrm{vir}$, diamonds are within $R_\mathrm{vir}$). 
Different colors refer to X-ray (red), kSZ including X-ray temperature bias (blue)
and ideal kSZ measurements (black). The middle panel shows the bias in form of the 
RMS of the velocity difference as function of the mass of the cluster and the right
panel shows the bias in form of the velocity difference.}
\label{fig:vel_scatter}
\end{figure*}

\section{Impact on observations} \label{sec:ksz}

It is outside the scope of this paper to include realistic observational 
effects through mimicking instruments or background. However, in this
section we want to evaluate the principal limits up to which accuracy
we can infer the peculiar velocity of clusters from observables, given 
the complex thermal and dynamical structure within galaxy clusters. We
also want to evaluate this in the light of different observables, which
intrinsically weight the radially varying signal differently. Especially we want
to investigate the influence of the radial dependent relative motion of the 
ICM in respect to the dark mater halo when measuring the ICM motion via 
kSZ effect or by measuring the shift of the iron line with detailed
X-ray spectroscopy.

\subsection{Spacial distribution of the observable signal}

To give an impression of the actual spacial distribution of the signal
we discuss the appearance of the different observables in the Appendix B1,
where Figure \ref{fig:sz_collage} a visualize the maps of different 
observable. To show, how far out from the cluster center
the the internal motions affect the kinematic signal, Figure \ref{fig:ksz_profile} shows the 
radially averaged, cumulative profile of the kinetic SZ effect. The different colors 
are the 8 different clusters, whereas the 3 different lines for each cluster are for three 
different projection directions of these four clusters. Clear to see that the internal 
motions are significantly influencing the kinematic signal up to a radius of between 0.3 and
0.5 times $R_\mathrm{vir}$.

\subsection{The observed bulk velocity}

To infer the mean bulk motion of the ICM from the kinetic SZ signal, one
has to assume, that the observed kinetic SZ signal, integrated over a certain
aperture $A$  
\begin{equation}
\left<w\right>_A = \int_A \frac{\sigma_t}{c} \int n_e v_r \mathrm{d}l \approx
                   \int_A \frac{\sigma_t \left<v_r\right>}{c} \int n_e \mathrm{d}l
\end{equation}
can be interpreted as the the signal of the mean (radial) velocity $\left<v_r\right>$ if the ICM, 
weighted with the mean surface mass density. The later can be obtained from the
integrated thermal SZ signal within the same aperture
\begin{equation}
\left<y\right>_A = \int_A \frac{k \sigma_T}{m_ec^2} \int n_e T \mathrm{d}l \approx 
                   \int_A \frac{k \sigma_T \left<T\right>}{m_ec^2} \int n_e \mathrm{d}l
\end{equation}
with the Boltzmann constant $k$, the Thomson cross-section $\sigma_T$, the speed of light $c$,
the mass of the electron $m_e$ and the temperature and density of the ICM $T$ and $n_e$.
Assuming that one can infer $\left<T\right>$ from X-ray observations, one can combine
these two measurements to obtain the mean velocity $\left<v_r\right>$ within the aperture $A$.

The aperture $A$ here can either be the beam in case the cluster could be resolved
by the observations or the area of the cluster itself in case it is not resolved.
In general, this means that the observed velocity $\left<v_r\right>$ within the aperture $A$
therefore differ from the mean, mass weighted velocity by the difference between
the mass weighted temperature (from $\left<y\right>_A$) and the temperature inferred from 
X-ray observations. For a more detailed discussion see \citet{2005MNRAS.356.1477D}.
The relation of these two temperatures is not trivial, as
the X-ray temperature is weighted by the emission measure and comes from a fit of an 
emission spectra usually assuming a single temperature. For the complex temperature
structure of galaxy clusters this can induce a significant bias 
\citep[see][]{2001ApJ...546..100M,2004MNRAS.354...10M,2006ApJ...640..710V}
This bias has to be either corrected using simulations,
or can be minimized by using spacial resolved X-ray temperature maps (which still
leaves the multi temperature structure along the line of sight) but allows to re-construct
the 3D temperature profile using iterative reconstruction methods \citep[see][and references therein]{2007MNRAS.382..397A}

Alternatively one can also use multi-temperature model 
\citep[see][]{2004A&A...413..415K,2012MNRAS.420.3545B}
to obtain the temperature distribution of the ICM.
For our purpose, we will investigate two extreme cases. In the most optimistic
case, we assume that this bias can be eliminated completely and we take the mass weighted
mean velocity along the line of sight (labeled as {\it ICM} in figure \ref{fig:vel_scatter}) 
within the aperture $A$. In the most pessimistic case, we include the temperature bias 
as inferred from the simulation (labeled as {\it kSZ} in figure \ref{fig:vel_scatter}).

In case the mean ICM velocity is inferred from X-ray spectroscopic measurements
(labeled as {\it X-ray} in figure \ref{fig:vel_scatter}), the observed velocity 
correspond to the mean, emission weighted ICM velocity, which due to the steep radial 
emission profile is biased towards the mean velocity within the cluster core. 

As the SZ measurements are not dependence on the distance, we use the extend
of the simulation (e.g. 300 Mpc/h) along the line of sight when measuring quantities
within the aperture $A$. As the X-ray signal is redshift and distance dependent, we use only
the local signal of the cluster when calculating the temperature bias and the 
emission weighted mean velocity within the same aperture $A$. We restrict ourselves
to explore 3 different apertures, namely the core (corresponding to $0.1\times{}R_{vir}$),
a region spanning $0.3\times{}R_{vir}$ and the region encompassing the virial radius.

In the left panel of figure \ref{fig:vel_scatter} we show a point by point comparison
of the difference of the observed velocity along the line of sight and the peculiar
velocity along the line of sight for all 9 combinations of measurements and apertures.
The middle panel of figure \ref{fig:vel_scatter} shows the root mean square (RMS) scatter
between the different observed velocities and the peculiar velocity of the cluster
and the right panel shows the mean between the different observed velocities and the 
peculiar velocity.
Clear to see that the X-ray based methods have larger scatter, as they emphasize the
relative motions of the ICM within the cores. Here it is also clearly visible, that
going to larger masses increases the difference, because more massive systems show
stronger sloshing motions as shown earlier. This effect dominates all measurements which
depend on small, central apertures. However, larger apertures show an opposite effect
when going to larger clusters. This is because within a more massive clusters, the larger
volume probed along the line of sight overcomes the increased bias within the central 
region.

We also note that the knowledge of the true temperature of the system is quite important.
Any bias in the measured temperature leads to a bias in the inferred kSZ signal, as
can be clearly be seen for the blue points in the left panel of figure \ref{fig:vel_scatter},
where we assumed the typical temperature bias as inferred from simulations. Note also
that especially in the systems with the largest velocity, the distribution of the individual
points clearly shows deviations from a linear relation which indicates that in these cases
calibrating a simple bias is not enough to recover the real signal.

\subsection{Stars as proxies for the bulk motions of clusters} \label{sec:proxies}

Finally we investigate alternatively to use the collissionless components of
the baryonic content of clusters as probe of the peculiar motion of the system.
To do so, we are using the highest resolutions simulation in our set. This allows us
to study in detail the stellar component of the simulations. Thereby, the larger, 
underlying resolution of this simulation allows us -- after subtracting all member 
galaxies -- to use the velocity distribution of the remaining stars to characterize
two, dynamically well-distinct stellar components within the simulated galaxy clusters. 
These differences in the dynamics
is then used to apply an unbinding procedure similar than applied when identifying the
member galaxies and lead to a spatial separation of the two components into stars belonging
to the central galaxy and stars belonging to the diffuse stellar component 
\citep[see][for details]{2010MNRAS.405.1544D}.
This then can be used to investigate if, and in case which, stellar 
component might be the best proxy to measure the real peculiar motion of the cluster.

\begin{figure}
\includegraphics[width=0.5\textwidth]{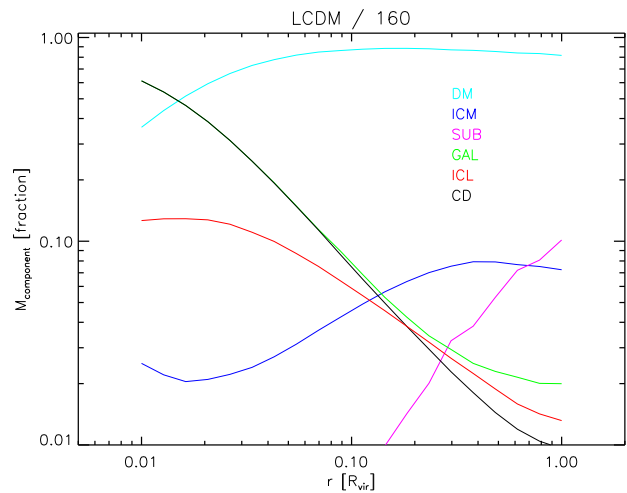}
\caption{Mass fraction of the different components of the clusters.
Shown is the average over the 24 most massive clusters in the high resolution,
{\it LCDM / 160} simulation. Here {\it SUB} devotes all material within identified
sub structures (excluding the main halo which also includes cD and ICL). {\it GAL}
devotes all stellar components of sub structures (but including the cD).}
\label{fig:cluster_composition}
\end{figure}

Although in this simulation, the most massive galaxy cluster is resolved with nearly
one million particles, we will present the results when averaging over the 24 most massive 
clusters within the {\it LCDM / 160} simulation, as the radial profiles obtained from
from individual clusters turn out to be still too noisy. The different stellar components 
have different radial distributions and reflect different amount of the baryonic component 
of galaxy clusters. Figure \ref{fig:cluster_composition} shows the relative mass 
fraction of the different components as function of radius. In the very center part, the 
stars from the central galaxy ({\it cD}) are dominating, outside of that region the dark 
matter ({\it DM}) is dominating by far. The {\it ICM} does not play an important role within the core and 
has a roughly constant fraction outside of it. In the very outer part, the total mass 
associated to the satellite galaxies ({\it SUB}) dominates, whereas the contribution
of the stellar mass of the sub-structures ({\it GAL}) as well as the diffuse stellar 
component ({\it ICL}) fall radially faster of than the dark matter.

\begin{figure}
\includegraphics[width=0.5\textwidth]{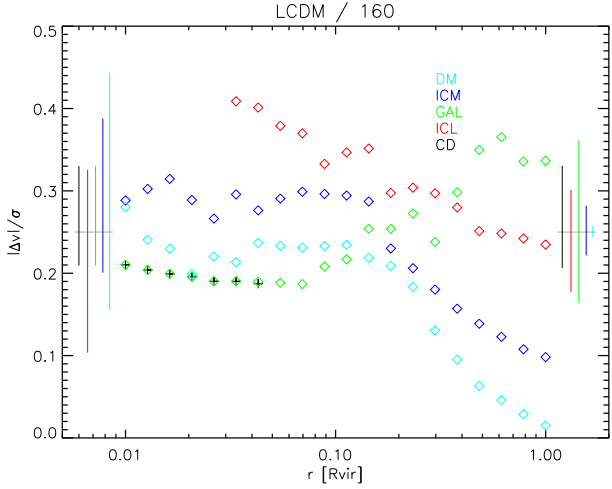}
\caption{Profiles (cumulative) of the relative motion of different components 
with respect to the underlying peculiar motion of the cluster. Shown is the 
again average over the 24 most massive clusters in the {\it LCDM / 160} simulation. 
Components are the same than in figure \ref{fig:cluster_composition}.
The error bars left and right to the curves show the 25/75 percentiles
typical at small (left) and large (right) radii.}
\label{fig:cluster_bulk motion}
\end{figure}

Finally, figure \ref{fig:cluster_bulk motion} shows how these different components
trace the underlying peculiar motion of the cluster as a whole. We again
show the result averaged over the 24 most massive clusters from the {\it LCDM / 160} 
simulation, where we normalized the velocity differences of the individual clusters
by their virial velocity dispersion obtained from equation (\ref{eq:sigma_v}) before
averaging, in the same spirit than done in the right panel of figure
\ref{fig:vel_prof_045}. In the central region, the relative motion of the 
dark matter component is slightly less than that one of the ICM component (around 25\% 
and 30\% of the virial velocity dispersion, respectively) but rapidly approaches the 
peculiar velocity of the halo with only a very small remaining bias at large distances,
originating from the small, mass weighed contribution of the baryonic component.
The best tracer of the peculiar motion of the cluster close to the center 
seems to be the stellar component of the central galaxy, which shows the
smallest relative velocity (ca. 20\% of the virial velocity dispersion).
As already noted earlier, the mean velocity of the galaxies is a quite bad
tracer of the peculiar motion of the cluster, especially when going to
the outer parts where more and more galaxies are contributing and the weight
of the central galaxy decreases. There the relative motion reaches
up to 35\% of the virial velocity dispersion. The diffuse stellar component performs
here much better (having only a relative motion of roughly 20\% of the virial velocity
dispersion). This is still significantly larger that the residual motions of the ICM
at this large distances and it remains to be investigated if that remaining signal is dominated
from the outer parts of the otherwise separately identified galaxies, which then might be 
associated to the diffuse component or if the velocity difference is driven by remaining
streams of already destroyed galaxies.
It has also to be stressed that the dispersion among the 24 individual is quite substantial,
as can be seen from the exemplary error bars displayed at small and large radii. 
Here we conclude that although having a large offset, the stellar component of the
central galaxy would give relatively the best proxy of the overall peculiar velocity of the cluster
at small, cluster centric distances, whereas at large distance the ICM measured
by the kinetic SZ effect will be still the best indicator.

\section{Application to pairwise velocity measures}

\begin{figure*}
\includegraphics[width=1.0\textwidth]{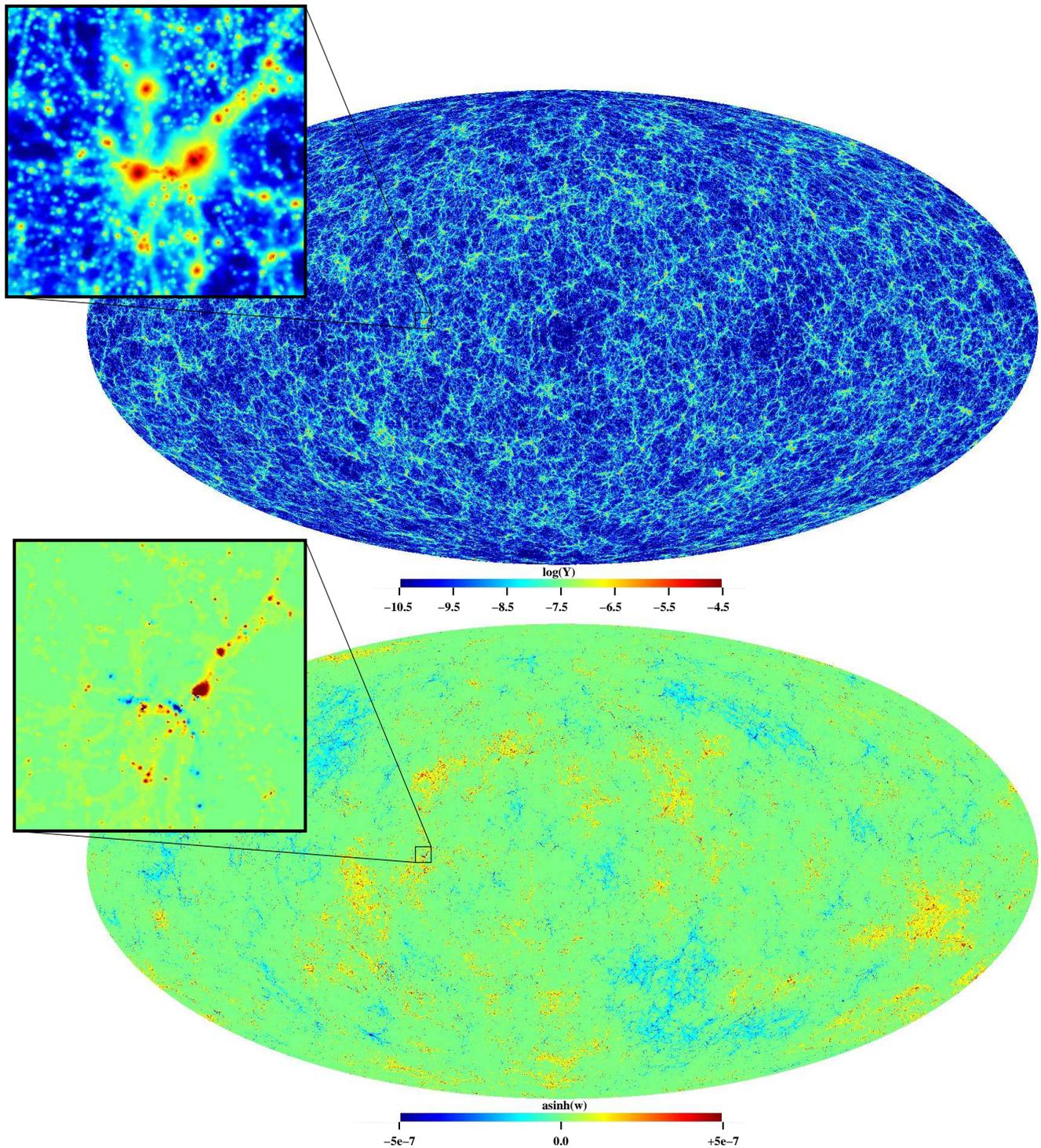}
\caption{Full sky projections of thermal (upper panel) and kinetic (lower panel) signal
taken exemplary from the last shell (e.g. $z=0.14-0.16$) integrated through the 
{\it WMAP5/1200} simulation. The inlays show a zoom onto one of the prominent structures
within that shell. Clear to see the large scale peculiar velocity patterns reflected in
the kinetic signal, which inside the structures get perturbed by the local motions within
the super cluster structure as well s internal motions within some of the clusters.} 
\label{fig:full_sky_maps}
\end{figure*}

To verify the findings presented in \citep{2012PhRvL.109d1101H} we where using
{\sc SMAC} \citep{dolag2005} to produced full sky maps from the simulations. Here we 
stack the simulation of the local universe ({\it LCDM/160}) into the large, 
cosmological box ({\it WMAP5/1200}). We constructed the full sky maps stacking
consecutive shells through the cosmological boxes taken at the evolution time
corresponding to the distance. Without replicating the box we thereby reached
a maximum distance of $z=0.16$. The according maps for the thermal and kinetic
SZ effect are displayed in figure \ref{fig:full_sky_maps}, which shows exemplary 
the last shell taken from the {\it WMAP5/1200} simulations. The maps are based
on a HEALPIx representation of the full sky {\citep{healpix} using a resolution
of $n_\mathrm{side} = 2048$ which leads to a pixel size roughly corresponding to 
the binning radius of the maps of the individual groups and clusters used in
\citet{2012PhRvL.109d1101H}. 

We then identified all galaxies (e.g. sub-halos) in the simulations. Applying
a cut in the maximum circular velocity ($v_\mathrm{max}$) we selected the most massive
galaxies, which represent the center of groups and clusters. Analog to 
\citet{2012PhRvL.109d1101H}, from the three dimensional position of these galaxies,
$\vec{r}_i$ we define the pairwise signal as
\begin{equation}
p_\mathrm{kSZ}(r) = - \frac{\sum_{i<j}(T_i - T_j)c_{ij}}{\sum_{i<j}c_{ij}^2}
\end{equation}
whereas $T_i$ and $T_j$ are the pixel values of the full sky map towards the position
of galaxies $i$ and $j$, respectively, and $c_{ij}$ is constructed from the three 
dimensional positions $\vec{r}_i$ and $\vec{r}_j$ according to
\begin{equation}
   c_{ij} = \vec{r}_{ij}\dot\frac{\vec{r}_i+\vec{r}_j}{2}.
\end{equation}

\begin{figure}
\includegraphics[width=0.5\textwidth]{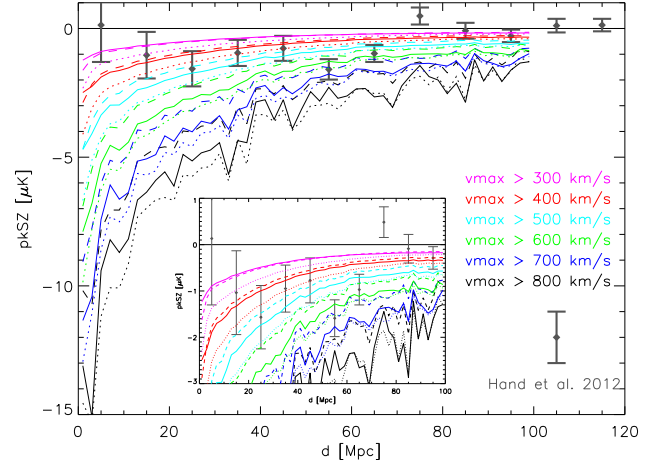}
\caption{Pairwise signal for different cuts in maximum circular velocity (e.g. Mass)
of galaxies from the simulations using the kinetic SZ map directly produced from the 
hydrodynamical simulations (solid lines) compared with the observational data 
points taken from \citet{2012PhRvL.109d1101H}. The dotted lines shows the 
estimated signal when using the combination of thermal SZ, temperature and bulk
velocity of the halo. The dashed line shows the same but calibrates the optical depth
to match the signal when using all halos. The inlay shows a zoom into the range where
the observational signal is detected.}
\label{fig:pairwise}
\end{figure}

Figure \ref{fig:pairwise} shows the result from this procedure compared to the
observational data points. When restricting to more massive systems, the signal
is, as expected, much stronger and extend to much larger distances. However,
the results obtained from the hydrodynamical
simulations are in good agreement with the signal seen in the observations. The
Luminosity cut applied to the observations should correspond to a limiting halo 
mass in the sample of $M_{200c} \approx 4.1\times10^{13} M_\odot$ \citep{2012PhRvL.109d1101H},
which corresponds roughly to a maximum, circular velocity of the simulated halos of 
$\approx 500 \mathrm{km/s}$. Note that the theoretical curves are based on the real, 
three dimensional position of the clusters/groups and the calculation does not
include observational effects like the contribution of the peculiar velocity of the 
individual clusters/group to their distance estimate or the averaging of the signal
at small scales due to the beam. Especially at the smallest bins it can be expected
that this would lead to a change of the signal.

\begin{figure}
\includegraphics[width=0.5\textwidth]{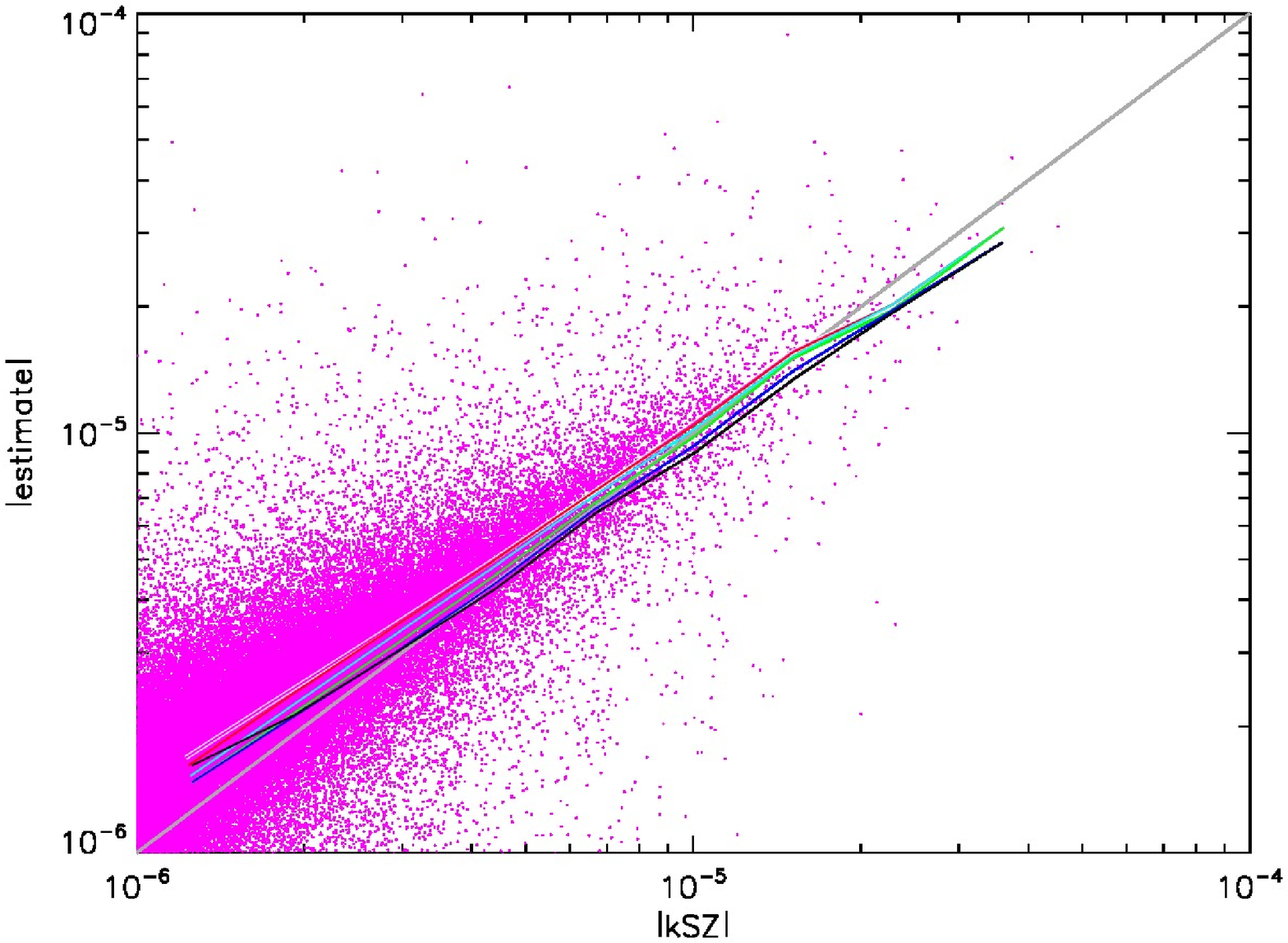}
\caption{Pixel by pixel comparison of the kinetic SZ signal obtained directly from the
hydrodynamical simulation with the estimated signal based on the bulk velocity of the halo.
Shown is the signal of all halos selected wit a circular velocity larger than $300 \mathrm{km/s}$.
Additional, the mean is shown for selecting different halos based on their circular velocity
as done is figure \ref{fig:pairwise}.}
\label{fig:pixelpixel}
\end{figure}

We also produced a mass weighed temperature map which then can be used when 
combining equation (5) and (6) to get an estimated kinetic SZ signal base on the   
halo bulk velocity and the thermal SZ signal
\begin{equation}
 \left<w\right>_A = v_\mathrm{halo} \frac{\left<y\right>_A}{\left<T\right>_A} \frac{m_e c}{k}.
\end{equation}
The resulting signal from this estimation can be compared pixel by pixel to the kinetic SZ signal
which was obtained directly from the hydrodynamical simulation. The result of that comparison for
all halos is shown in figure \ref{fig:pixelpixel}. Clear to see that there is not only a large 
scatter but also systematic trend which is related by approximating the optical depth through the
mean temperature and the mean thermal SZ measured in each pixel. The expected pairwise signal from 
this approximation is shown as dotted line in figure \ref{fig:pairwise}. The bias induced by
approximating the optical depth of the individual systems by combining their thermal SZ signal with the 
mean temperature of these systems can in principal be calibrated based on the hydrodynamical simulations
itself. We find that a reduction of 30\% of the estimated optical depth leads to a very good 
agreement between the true and the estimated signal, especially when looking at the signal
obtained when including all halos, as can be seen by the dashed lines in figure \ref{fig:pairwise},
which falls basically on top of the signal obtained from the real kinetic SZ signal extracted
from the maps. However, when going to more massive systems it can be seen that the true kinetic SZ signal is
significantly larger than the one estimated from the bulk velocity of the halo, again indicating
that in massive systems internal motions of the ICM in respect to the bulk motion of the halo itself 
are contributing significantly to the expected kinetic SZ signal.

%##################################################################################
%########################## Conclusions ###########################################
%##################################################################################

\section{Conclusions} \label{sec:conc}

We used a large set of cosmological, hydrodynamical simulations, which cover very large 
cosmological volume, hosting a large number of rich clusters of galaxies, as well
as moderate volumes where the internal structures of individual galaxy clusters can be
resolved with very high resolution to investigate, how the presence of baryons and their
associated physical processes like cooling and star-formation are affecting the
systematic difference between mass averaged velocities of dark matter and the ICM 
inside a cluster and how that is reflected in observables like the kinematic SZ effect.

Our main findings are:

\begin{itemize}
\item The peculiar velocity of galaxy clusters are, as expected, independent of mass 
and redshift following a Maxwell distribution. However, being able to probe very large 
cosmological volumes, we could demonstrate that the characteristic velocity of this
distribution is not a fixed value but depends on the local, large scale over-density
within a sphere of 20 Mpc ($\delta_{20\mathrm{Mpc}}$) as
\begin{equation}
   \sigma_v \; [\mathrm{km}/\mathrm{s}] \; = 185 \times \mathrm{exp}\left(\delta_{20\mathrm{Mpc}}/10\right)
\end{equation}
and thereby typical peculiar velocities (which are $\approx$ 410 km/s for the most recent
cosmological parameters) increase by a factor of 2 within super cluster regions.

\item Also in merging systems the relative velocity of the clusters is strongly
increase and typically exceeds values of 1000 km/s in the moment when the virial sizes of the
two systems touch. It has to be also noted that even in the larges box of or simulation
campaign ({\it WMAP5 / 1200}) all cluster pairs  with distances less than
three times their virial size are approaching each other.

\item Relative ICM motions in the center of galaxy clusters can in some cases exceed the
peculiar motion of the cluster as a whole. We find relative motions (so called sloshing) 
of the ICM up to 1000 km/s, in good agreement with observations. This sloshing motions
are scaling with the virial velocity dispersion of the clusters and typically 25\% of the
clusters show sloshing motions of 20\% of the virial velocity dispersion within their 
core which decrease to 9\% averaged over the virial region. The relative motions
of the core compared to the peculiar velocity thereby are completely uncorrelated.

\item When measuring the peculiar motion of the galaxy cluster by combining the 
kinematic and thermal SZ measurement, the additional bias between the mass weighted
temperature (entering the thermal SZ signal) and the temperature inferred from X-ray
observations (which is in addition needed) further increases the scatter between
the inferred velocity and the true, peculiar velocity. The combination of the 
dependence of the relative motions of the ICM with cluster mass and the dependence of
the observed signal with the size of the observational aperture leads to a mass
dependence of this scatter. Whereas for less massive clusters (e.g. $10^{14} M_\odot/h$), 
the induced scatter is of order 60-90 km/s dependent on the aperture and the
size of the temperature bias, for very massive systems (e.g. $10^{14} M_\odot/h$),
the scatter is between 50 km/s and 200 km/s, depending mainly on the size of the
aperture. In case the signal is inferred from X-ray data only (e.g. using high 
resolution microcaloimeters), the scatter for small apertures is even significantly
larger. 

\item When looking at the stellar components of the cluster it is important to note
that the mean velocity of the galaxy population even when averaging within the
virial region is a very bad indicator of the peculiar velocity of the system.
Interestingly, although obtained only from the central part of the cluster, 
the stellar component of the central galaxy gives a much better indicator
of the peculiar motion of the cluster, however the relative difference in that
case is still a factor of two worse than the mean velocity obtained from 
the ICM within the virial region.
 
\item The pairwise velocity signal as measured in \citet{2012PhRvL.109d1101H}
is in qualitative agreement with the expectations from hydrodynamical simulations,
although it strongly depends on the mass cut applied to the selected clusters/groups.
There is a significant (30\%) bias introduced by the estimation of the optical depth
based on temperature and thermal SZ signal. Additional, when restricting to massive
system to increase the signal, internal velocities of the ICM are giving a significant
extra contribution to the observed kinetic SZ effect.

\end{itemize}

Given the improved sensitivity of current instruments it might be expected that 
the kinetic SZ signal in massive clusters will be detected soon. However, it 
has to be kept in mind that, as we clearly demonstrated, the internal motions 
of the ICM within the central part in respect to the galaxy cluster as whole
is more pronounced in massive systems, where such internal motions by far
exceed the expected bulk motion, which leads to an unavoidable systematical error 
in the inferred peculiar velocities of the halos which is connected to the
relative motion between the dark matter and the ICM. 
This will dominate the signal unless one
integrates over relatively large area. However, when integrating over larger
area, understanding and correcting for the the bias between mass weighted and 
measured (via X-ray) temperature gets more crucial, as one averages over a larger
range in temperature structures within the ICM. Measurements of the ICM velocity
using X-ray spectra will be even more affected, as they give additional weight
to the central regions. Therefore we conclude that measuring the velocity
of the ICM in the central regions of galaxy clusters will give very interesting 
insights into the dynamics of galaxy clusters, however one has to be carefully
when associating the peculiar velocity of galaxy clusters with such measurements.

%##################################################################################
%########################## Acknowledgments / Bibliography #######################
%#################################################################################

\section*{acknowledgments}
We are indebted to L. Moscardini for useful discussions and comments.
Computations have been performed at the “Leibniz-Rechenzentrum” with CPU time 
assigned to the Project “h0073”. K.D. acknowledges the support by the “HPC-Europa2
Transnational Access program”, the DFG Priority Programme 1177 and additional 
support by the DFG Cluster of Excellence "Origin and Structure of the Universe". 

\bibliographystyle{mn2e}
\bibliography{master,master3}

\begin{thebibliography}{}

\bibitem[\protect\citeauthoryear{{Ameglio}, {Borgani}, {Pierpaoli} \&
  {Dolag}}{{Ameglio} et~al.}{2007}]{2007MNRAS.382..397A}
{Ameglio} S.,  {Borgani} S.,  {Pierpaoli} E.,    {Dolag} K.,  2007, \mnras,
  382, 397

\bibitem[\protect\citeauthoryear{{Ascasibar} \& {Markevitch}}{{Ascasibar} \&
  {Markevitch}}{2006}]{2006ApJ...650..102A}
{Ascasibar} Y.,  {Markevitch} M.,  2006, \apj, 650, 102

\bibitem[\protect\citeauthoryear{{Bahcall}, {Gramann} \& {Cen}}{{Bahcall}
  et~al.}{1994}]{1994ApJ...436...23B}
{Bahcall} N.~A.,  {Gramann} M.,    {Cen} R.,  1994, \apj, 436, 23

\bibitem[\protect\citeauthoryear{{Biffi}, {Dolag}, {B{\"o}hringer} \&
  {Lemson}}{{Biffi} et~al.}{2012}]{2012MNRAS.420.3545B}
{Biffi} V.,  {Dolag} K.,  {B{\"o}hringer} H.,    {Lemson} G.,  2012, \mnras,
  420, 3545

\bibitem[\protect\citeauthoryear{{Bryan} \& {Norman}}{{Bryan} \&
  {Norman}}{1998}]{bryan1998}
{Bryan} G.~L.,  {Norman} M.~L.,  1998, \apj, 495, 80

\bibitem[\protect\citeauthoryear{{Bulbul}, {Smith}, {Foster}, {Cottam},
  {Loewenstein}, {Mushotzky} \& {Shafer}}{{Bulbul}
  et~al.}{2012}]{2012ApJ...747...32B}
{Bulbul} G.~E.,  {Smith} R.~K.,  {Foster} A.,  {Cottam} J.,  {Loewenstein} M.,
  {Mushotzky} R.,    {Shafer} R.,  2012, \apj, 747, 32

\bibitem[\protect\citeauthoryear{{Churazov}, {Br{\"u}ggen}, {Kaiser},
  {B{\"o}hringer} \& {Forman}}{{Churazov} et~al.}{2001}]{2001ApJ...554..261C}
{Churazov} E.,  {Br{\"u}ggen} M.,  {Kaiser} C.~R.,  {B{\"o}hringer} H.,
  {Forman} W.,  2001, \apj, 554, 261

\bibitem[\protect\citeauthoryear{{Churazov}, {Vikhlinin}, {Zhuravleva},
  {Schekochihin}, {Parrish}, {Sunyaev}, {Forman}, {B{\"o}hringer} \&
  {Randall}}{{Churazov} et~al.}{2012}]{2012MNRAS.421.1123C}
{Churazov} E.,  {Vikhlinin} A.,  {Zhuravleva} I.,  {Schekochihin} A.,
  {Parrish} I.,  {Sunyaev} R.,  {Forman} W.,  {B{\"o}hringer} H.,    {Randall}
  S.,  2012, \mnras, 421, 1123

\bibitem[\protect\citeauthoryear{{De Boni}, {Dolag}, {Ettori}, {Moscardini},
  {Pettorino} \& {Baccigalupi}}{{De Boni} et~al.}{2011}]{2011MNRAS.415.2758D}
{De Boni} C.,  {Dolag} K.,  {Ettori} S.,  {Moscardini} L.,  {Pettorino} V.,
  {Baccigalupi} C.,  2011, \mnras, 415, 2758

\bibitem[\protect\citeauthoryear{{De Boni}, {Ettori}, {Dolag} \&
  {Moscardini}}{{De Boni} et~al.}{2012}]{2012arXiv1205.3163D}
{De Boni} C.,  {Ettori} S.,  {Dolag} K.,    {Moscardini} L.,  2012, ArXiv
  e-prints

\bibitem[\protect\citeauthoryear{{Diaferio}, {Borgani}, {Moscardini},
  {Murante}, {Dolag}, {Springel}, {Tormen}, {Tornatore} \& {Tozzi}}{{Diaferio}
  et~al.}{2005}]{2005MNRAS.356.1477D}
{Diaferio} A.,  {Borgani} S.,  {Moscardini} L.,  {Murante} G.,  {Dolag} K.,
  {Springel} V.,  {Tormen} G.,  {Tornatore} L.,    {Tozzi} P.,  2005, \mnras,
  356, 1477

\bibitem[\protect\citeauthoryear{{Diaferio}, {Sunyaev} \& {Nusser}}{{Diaferio}
  et~al.}{2000}]{2000ApJ...533L..71D}
{Diaferio} A.,  {Sunyaev} R.~A.,    {Nusser} A.,  2000, \apjl, 533, L71

\bibitem[\protect\citeauthoryear{{Diego}, {Mazzotta} \& {Silk}}{{Diego}
  et~al.}{2003}]{2003ApJ...597L...1D}
{Diego} J.~M.,  {Mazzotta} P.,    {Silk} J.,  2003, \apjl, 597, L1

\bibitem[\protect\citeauthoryear{{Dolag}, {Borgani}, {Murante} \&
  {Springel}}{{Dolag} et~al.}{2009}]{2009MNRAS.399..497D}
{Dolag} K.,  {Borgani} S.,  {Murante} G.,    {Springel} V.,  2009, \mnras, 399,
  497

\bibitem[\protect\citeauthoryear{{Dolag}, {Hansen}, {Roncarelli} \&
  {Moscardini}}{{Dolag} et~al.}{2005}]{dolag2005}
{Dolag} K.,  {Hansen} F.~K.,  {Roncarelli} M.,    {Moscardini} L.,  2005,
  \mnras, 363, 29

\bibitem[\protect\citeauthoryear{{Dolag}, {Murante} \& {Borgani}}{{Dolag}
  et~al.}{2010}]{2010MNRAS.405.1544D}
{Dolag} K.,  {Murante} G.,    {Borgani} S.,  2010, \mnras, 405, 1544

\bibitem[\protect\citeauthoryear{{Dolag}, {Reinecke}, {Gheller} \&
  {Imboden}}{{Dolag} et~al.}{2008}]{Dolag08}
{Dolag} K.,  {Reinecke} M.,  {Gheller} C.,    {Imboden} S.,  2008, NJoP,
  submitted

\bibitem[\protect\citeauthoryear{{Dolag}, {Vazza}, {Brunetti} \&
  {Tormen}}{{Dolag} et~al.}{2005}]{2005MNRAS.364..753D}
{Dolag} K.,  {Vazza} F.,  {Brunetti} G.,    {Tormen} G.,  2005, \mnras, 364,
  753

\bibitem[\protect\citeauthoryear{{Fedeli}, {Dolag} \& {Moscardini}}{{Fedeli}
  et~al.}{2012}]{2012MNRAS.419.1588F}
{Fedeli} C.,  {Dolag} K.,    {Moscardini} L.,  2012, \mnras, 419, 1588

\bibitem[\protect\citeauthoryear{{Frenk}, {White}, {Bode}, {Bond}, {Bryan},
  {Cen}, {Couchman}, {Evrard}, {Gnedin}, {Jenkins}, {Khokhlov} \&
  {Klypin}}{{Frenk} et~al.}{1999}]{1999ApJ...525..554F}
{Frenk} C.~S.,  {White} S.~D.~M.,  {Bode} P.,  {Bond} J.~R.,  {Bryan} G.~L.,
  {Cen} R.,  {Couchman} H.~M.~P.,  {Evrard} A.~E.,  {Gnedin} N.,  {Jenkins} A.,
   {Khokhlov} A.~M.,    {Klypin} A.,  1999, \apj, 525, 554

\bibitem[\protect\citeauthoryear{G\'orski, Hivon \& Wandelt}{G\'orski
  et~al.}{1998}]{healpix}
G\'orski K.~M.,  Hivon E.,    Wandelt B.~D.,  1998, `Analysis Issues for Large
  CMB Data Sets', 1998, eds A. J. Banday,R. K. Sheth and L. Da Costa, ESO,
  Printpartners Ipskamp, NL, pp.37-42 (astro-ph/9812350); Healpix HOMEPAGE:
  http://www.eso.org/science/healpix/

\bibitem[\protect\citeauthoryear{{Grossi}, {Verde}, {Carbone}, {Dolag},
  {Branchini}, {Iannuzzi}, {Matarrese} \& {Moscardini}}{{Grossi}
  et~al.}{2009}]{2009MNRAS.398..321G}
{Grossi} M.,  {Verde} L.,  {Carbone} C.,  {Dolag} K.,  {Branchini} E.,
  {Iannuzzi} F.,  {Matarrese} S.,    {Moscardini} L.,  2009, \mnras, 398, 321

\bibitem[\protect\citeauthoryear{{Haardt} \& {Madau}}{{Haardt} \&
  {Madau}}{1996}]{1996ApJ...461...20H}
{Haardt} F.,  {Madau} P.,  1996, \apj, 461, 20

\bibitem[\protect\citeauthoryear{{Hand}, {Addison}, {Aubourg}, {Battaglia},
  {Battistelli}, {Bizyaev}, {Bond}, {Brewington}, {Brinkmann}, {Brown}, {Das},
  {Dawson}, {Devlin}, {Dunkley} \& {Dunner}}{{Hand}
  et~al.}{2012}]{2012PhRvL.109d1101H}
{Hand} N.,  {Addison} G.~E.,  {Aubourg} E.,  {Battaglia} N.,  {Battistelli}
  E.~S.,  {Bizyaev} D.,  {Bond} J.~R.,  {Brewington} H.,  {Brinkmann} J.,
  {Brown} B.~R.,  {Das} S.,  {Dawson} K.~S.,  {Devlin} M.~J.,  {Dunkley} J.,
  {Dunner} R. e.~a.,  2012, Physical Review Letters, 109, 041101

\bibitem[\protect\citeauthoryear{{Hoffman} \& {Ribak}}{{Hoffman} \&
  {Ribak}}{1991}]{Hoffman1991}
{Hoffman} Y.,  {Ribak} E.,  1991, \apj, 380, L5

\bibitem[\protect\citeauthoryear{{Iapichino} \& {Niemeyer}}{{Iapichino} \&
  {Niemeyer}}{2008}]{2008MNRAS.388.1089I}
{Iapichino} L.,  {Niemeyer} J.~C.,  2008, \mnras, 388, 1089

\bibitem[\protect\citeauthoryear{{Inogamov} \& {Sunyaev}}{{Inogamov} \&
  {Sunyaev}}{2003}]{2003AstL...29..791I}
{Inogamov} N.~A.,  {Sunyaev} R.~A.,  2003, Astronomy Letters, 29, 791

\bibitem[\protect\citeauthoryear{{Kaastra}, {Tamura}, {Peterson}, {Bleeker},
  {Ferrigno}, {Kahn}, {Paerels}, {Piffaretti}, {Branduardi-Raymont} \&
  {B{\"o}hringer}}{{Kaastra} et~al.}{2004}]{2004A&A...413..415K}
{Kaastra} J.~S.,  {Tamura} T.,  {Peterson} J.~R.,  {Bleeker} J.~A.~M.,
  {Ferrigno} C.,  {Kahn} S.~M.,  {Paerels} F.~B.~S.,  {Piffaretti} R.,
  {Branduardi-Raymont} G.,    {B{\"o}hringer} H.,  2004, \aap, 413, 415

\bibitem[\protect\citeauthoryear{{Keisler} \& {Schmidt}}{{Keisler} \&
  {Schmidt}}{2012}]{2012arXiv1211.0668K}
{Keisler} R.,  {Schmidt} F.,  2012, ArXiv e-prints

\bibitem[\protect\citeauthoryear{{Kolatt}, {Dekel}, {Ganon} \&
  {Willick}}{{Kolatt} et~al.}{1996}]{Kolatt:1996}
{Kolatt} T.,  {Dekel} A.,  {Ganon} G.,    {Willick} J.~A.,  1996, \apj, 458,
  419

\bibitem[\protect\citeauthoryear{{Komatsu}, {Dunkley}, {Nolta}, {Bennett},
  {Gold}, {Hinshaw}, {Jarosik}, {Larson}, {Limon}, {Page}, {Spergel},
  {Halpern}, {Hill}, {Kogut}, {Meyer}, {Tucker}, {Weiland}, {Wollack} \&
  {Wright}}{{Komatsu} et~al.}{2009}]{2009ApJS..180..330K}
{Komatsu} E.,  {Dunkley} J.,  {Nolta} M.~R.,  {Bennett} C.~L.,  {Gold} B.,
  {Hinshaw} G.,  {Jarosik} N.,  {Larson} D.,  {Limon} M.,  {Page} L.,
  {Spergel} D.~N.,  {Halpern} M.,  {Hill} R.~S.,  {Kogut} A.,  {Meyer} S.~S.,
  {Tucker} G.~S.,  {Weiland} J.~L.,  {Wollack} E.,    {Wright} E.~L.,  2009,
  \apjs, 180, 330

\bibitem[\protect\citeauthoryear{{Markevitch}, {Gonzalez}, {Clowe},
  {Vikhlinin}, {Forman}, {Jones}, {Murray} \& {Tucker}}{{Markevitch}
  et~al.}{2004}]{2004ApJ...606..819M}
{Markevitch} M.,  {Gonzalez} A.~H.,  {Clowe} D.,  {Vikhlinin} A.,  {Forman} W.,
   {Jones} C.,  {Murray} S.,    {Tucker} W.,  2004, \apj, 606, 819

\bibitem[\protect\citeauthoryear{{Markevitch} \& {Vikhlinin}}{{Markevitch} \&
  {Vikhlinin}}{2007}]{2007PhR...443....1M}
{Markevitch} M.,  {Vikhlinin} A.,  2007, \physrep, 443, 1

\bibitem[\protect\citeauthoryear{{Mathiesen} \& {Evrard}}{{Mathiesen} \&
  {Evrard}}{2001}]{2001ApJ...546..100M}
{Mathiesen} B.~F.,  {Evrard} A.~E.,  2001, \apj, 546, 100

\bibitem[\protect\citeauthoryear{{Mathis}, {Lemson}, {Springel}, {Kauffmann},
  {White}, {Eldar} \& {Dekel}}{{Mathis} et~al.}{2002}]{Mathis:2002}
{Mathis} H.,  {Lemson} G.,  {Springel} V.,  {Kauffmann} G.,  {White} S.~D.~M.,
  {Eldar} A.,    {Dekel} A.,  2002, \mnras, 333, 739

\bibitem[\protect\citeauthoryear{{Mazzotta}, {Rasia}, {Moscardini} \&
  {Tormen}}{{Mazzotta} et~al.}{2004}]{2004MNRAS.354...10M}
{Mazzotta} P.,  {Rasia} E.,  {Moscardini} L.,    {Tormen} G.,  2004, \mnras,
  354, 10

\bibitem[\protect\citeauthoryear{{Nagai}, {Kravtsov} \& {Kosowsky}}{{Nagai}
  et~al.}{2003}]{2003ApJ...587..524N}
{Nagai} D.,  {Kravtsov} A.~V.,    {Kosowsky} A.,  2003, \apj, 587, 524

\bibitem[\protect\citeauthoryear{{Paul}, {Iapichino}, {Miniati}, {Bagchi} \&
  {Mannheim}}{{Paul} et~al.}{2011}]{2011ApJ...726...17P}
{Paul} S.,  {Iapichino} L.,  {Miniati} F.,  {Bagchi} J.,    {Mannheim} K.,
  2011, \apj, 726, 17

\bibitem[\protect\citeauthoryear{{Planck Collaboration}}{{Planck
  Collaboration}}{2012}]{2012arXiv1204.2743P}
{Planck Collaboration} 2012, ArXiv e-prints

\bibitem[\protect\citeauthoryear{{Rasia}, {Tormen} \& {Moscardini}}{{Rasia}
  et~al.}{2004}]{rasia2004}
{Rasia} E.,  {Tormen} G.,    {Moscardini} L.,  2004, \mnras, 351, 237

\bibitem[\protect\citeauthoryear{{Reid}, {Verde}, {Dolag}, {Matarrese} \&
  {Moscardini}}{{Reid} et~al.}{2010}]{2010JCAP...07..013R}
{Reid} B.~A.,  {Verde} L.,  {Dolag} K.,  {Matarrese} S.,    {Moscardini} L.,
  2010, \jcap, 7, 13

\bibitem[\protect\citeauthoryear{{Roediger}, {Br{\"u}ggen}, {Simionescu},
  {B{\"o}hringer}, {Churazov} \& {Forman}}{{Roediger}
  et~al.}{2011}]{2011MNRAS.413.2057R}
{Roediger} E.,  {Br{\"u}ggen} M.,  {Simionescu} A.,  {B{\"o}hringer} H.,
  {Churazov} E.,    {Forman} W.~R.,  2011, \mnras, 413, 2057

\bibitem[\protect\citeauthoryear{{Roncarelli}, {Moscardini}, {Branchini},
  {Dolag}, {Grossi}, {Iannuzzi} \& {Matarrese}}{{Roncarelli}
  et~al.}{2010}]{2010MNRAS.402..923R}
{Roncarelli} M.,  {Moscardini} L.,  {Branchini} E.,  {Dolag} K.,  {Grossi} M.,
  {Iannuzzi} F.,    {Matarrese} S.,  2010, \mnras, 402, 923

\bibitem[\protect\citeauthoryear{{Salpeter}}{{Salpeter}}{1955}]{1955ApJ...121..161S}
{Salpeter} E.~E.,  1955, \apj, 121, 161

\bibitem[\protect\citeauthoryear{{Sanders}, {Fabian} \& {Smith}}{{Sanders}
  et~al.}{2011}]{2011MNRAS.410.1797S}
{Sanders} J.~S.,  {Fabian} A.~C.,    {Smith} R.~K.,  2011, \mnras, 410, 1797

\bibitem[\protect\citeauthoryear{{Schuecker}, {Finoguenov}, {Miniati},
  {B{\"o}hringer} \& {Briel}}{{Schuecker} et~al.}{2004}]{2004A&A...426..387S}
{Schuecker} P.,  {Finoguenov} A.,  {Miniati} F.,  {B{\"o}hringer} H.,
  {Briel} U.~G.,  2004, \aap, 426, 387

\bibitem[\protect\citeauthoryear{{Simionescu}, {Werner}, {Urban}, {Allen},
  {Fabian}, {Sanders}, {Mantz}, {Nulsen} \& {Takei}}{{Simionescu}
  et~al.}{2012}]{2012arXiv1208.2990S}
{Simionescu} A.,  {Werner} N.,  {Urban} O.,  {Allen} S.~W.,  {Fabian} A.~C.,
  {Sanders} J.~S.,  {Mantz} A.,  {Nulsen} P.~E.~J.,    {Takei} Y.,  2012, ArXiv
  e-prints

\bibitem[\protect\citeauthoryear{{Spergel}, {Bean}, {Dor{\'e}}, {Nolta},
  {Bennett}, {Dunkley}, {Hinshaw}, {Jarosik}, {Komatsu}, {Page}, {Peiris},
  {Verde} \& Co-Authors}{{Spergel} et~al.}{2007}]{2007ApJS..170..377S}
{Spergel} D.~N.,  {Bean} R.,  {Dor{\'e}} O.,  {Nolta} M.~R.,  {Bennett} C.~L.,
  {Dunkley} J.,  {Hinshaw} G.,  {Jarosik} N.,  {Komatsu} E.,  {Page} L.,
  {Peiris} H.~V.,  {Verde} L.,    Co-Authors .,  2007, \apjs, 170, 377

\bibitem[\protect\citeauthoryear{{Springel}}{{Springel}}{2005}]{2005MNRAS.364.1105S}
{Springel} V.,  2005, \mnras, 364, 1105

\bibitem[\protect\citeauthoryear{{Springel} \& {Hernquist}}{{Springel} \&
  {Hernquist}}{2002}]{2002MNRAS.333..649S}
{Springel} V.,  {Hernquist} L.,  2002, \mnras, 333, 649

\bibitem[\protect\citeauthoryear{{Springel} \& {Hernquist}}{{Springel} \&
  {Hernquist}}{2003}]{2003MNRAS.339..289S}
{Springel} V.,  {Hernquist} L.,  2003, \mnras, 339, 289

\bibitem[\protect\citeauthoryear{{Springel}, {White}, {Tormen} \&
  {Kauffmann}}{{Springel} et~al.}{2001}]{2001MNRAS.328..726S}
{Springel} V.,  {White} S.~D.~M.,  {Tormen} G.,    {Kauffmann} G.,  2001,
  \mnras, 328, 726

\bibitem[\protect\citeauthoryear{{Springel}, {Yoshida} \& {White}}{{Springel}
  et~al.}{2001}]{2001ApJ...549..681S}
{Springel} V.,  {Yoshida} N.,    {White} S.~D.~M.,  2001, New Astronomy, 6, 79

\bibitem[\protect\citeauthoryear{{Sunyaev}, {Norman} \& {Bryan}}{{Sunyaev}
  et~al.}{2003}]{2003AstL...29..783S}
{Sunyaev} R.~A.,  {Norman} M.~L.,    {Bryan} G.~L.,  2003, Astronomy Letters,
  29, 783

\bibitem[\protect\citeauthoryear{{Sunyaev} \& {Zeldovich}}{{Sunyaev} \&
  {Zeldovich}}{1980}]{1980MNRAS.190..413S}
{Sunyaev} R.~A.,  {Zeldovich} I.~B.,  1980, \mnras, 190, 413

\bibitem[\protect\citeauthoryear{{Tornatore}, {Borgani}, {Dolag} \&
  {Matteucci}}{{Tornatore} et~al.}{2007}]{2007MNRAS.382.1050T}
{Tornatore} L.,  {Borgani} S.,  {Dolag} K.,    {Matteucci} F.,  2007, \mnras,
  382, 1050

\bibitem[\protect\citeauthoryear{{Tornatore}, {Borgani}, {Matteucci}, {Recchi}
  \& {Tozzi}}{{Tornatore} et~al.}{2004}]{2004MNRAS.349L..19T}
{Tornatore} L.,  {Borgani} S.,  {Matteucci} F.,  {Recchi} S.,    {Tozzi} P.,
  2004, \mnras, 349, L19

\bibitem[\protect\citeauthoryear{{Vazza}, {Brunetti}, {Gheller}, {Brunino} \&
  {Br{\"u}ggen}}{{Vazza} et~al.}{2011}]{2011A&A...529A..17V}
{Vazza} F.,  {Brunetti} G.,  {Gheller} C.,  {Brunino} R.,    {Br{\"u}ggen} M.,
  2011, \aap, 529, A17+

\bibitem[\protect\citeauthoryear{{Vazza}, {Brunetti}, {Kritsuk}, {Wagner},
  {Gheller} \& {Norman}}{{Vazza} et~al.}{2009}]{2009A&A...504...33V}
{Vazza} F.,  {Brunetti} G.,  {Kritsuk} A.,  {Wagner} R.,  {Gheller} C.,
  {Norman} M.,  2009, \aap, 504, 33

\bibitem[\protect\citeauthoryear{{Vazza}, {Tormen}, {Cassano}, {Brunetti} \&
  {Dolag}}{{Vazza} et~al.}{2006}]{2006MNRAS.369L..14V}
{Vazza} F.,  {Tormen} G.,  {Cassano} R.,  {Brunetti} G.,    {Dolag} K.,  2006,
  \mnras, 369, L14

\bibitem[\protect\citeauthoryear{{Vikhlinin}}{{Vikhlinin}}{2006}]{2006ApJ...640..710V}
{Vikhlinin} A.,  2006, \apj, 640, 710

\bibitem[\protect\citeauthoryear{{Zitrin}, {Bartelmann}, {Umetsu}, {Oguri} \&
  {Broadhurst}}{{Zitrin} et~al.}{2012}]{2012arXiv1208.1766Z}
{Zitrin} A.,  {Bartelmann} M.,  {Umetsu} K.,  {Oguri} M.,    {Broadhurst} T.,
  2012, ArXiv e-prints

\bibitem[\protect\citeauthoryear{{ZuHone}, {Markevitch} \& {Johnson}}{{ZuHone}
  et~al.}{2010}]{2010ApJ...717..908Z}
{ZuHone} J.~A.,  {Markevitch} M.,    {Johnson} R.~E.,  2010, \apj, 717, 908

\end{thebibliography}

\appendix

\section{Peculiar velocity of halos in DM and csf simulations}

The distribution of the overall peculiar motions of galaxy clusters reflect 
the large scale velocity field, which still probes 
the linear regime of structure formation. Therefore we do not expect the presence 
of baryonic components and their associated physical processes (like cooling, 
star-formation and stellar feedback) to influence the overall peculiar velocities
of galaxy clusters. This is also demonstrated in figure \ref{fig:halo_vel_hist_mass_gas},
which shows (as in the left panel of figure \ref{fig:halo_vel_hist}) the distribution 
of cluster peculiar velocities for different cut in cluster mass from the large cosmological
box {\it WMAP5 / 1200}, comparing the simulation including cooling and star-formation (solid line)
with the dark matter only simulation of he same cosmological box (dashed lines).

\begin{figure}
\includegraphics[width=0.5\textwidth]{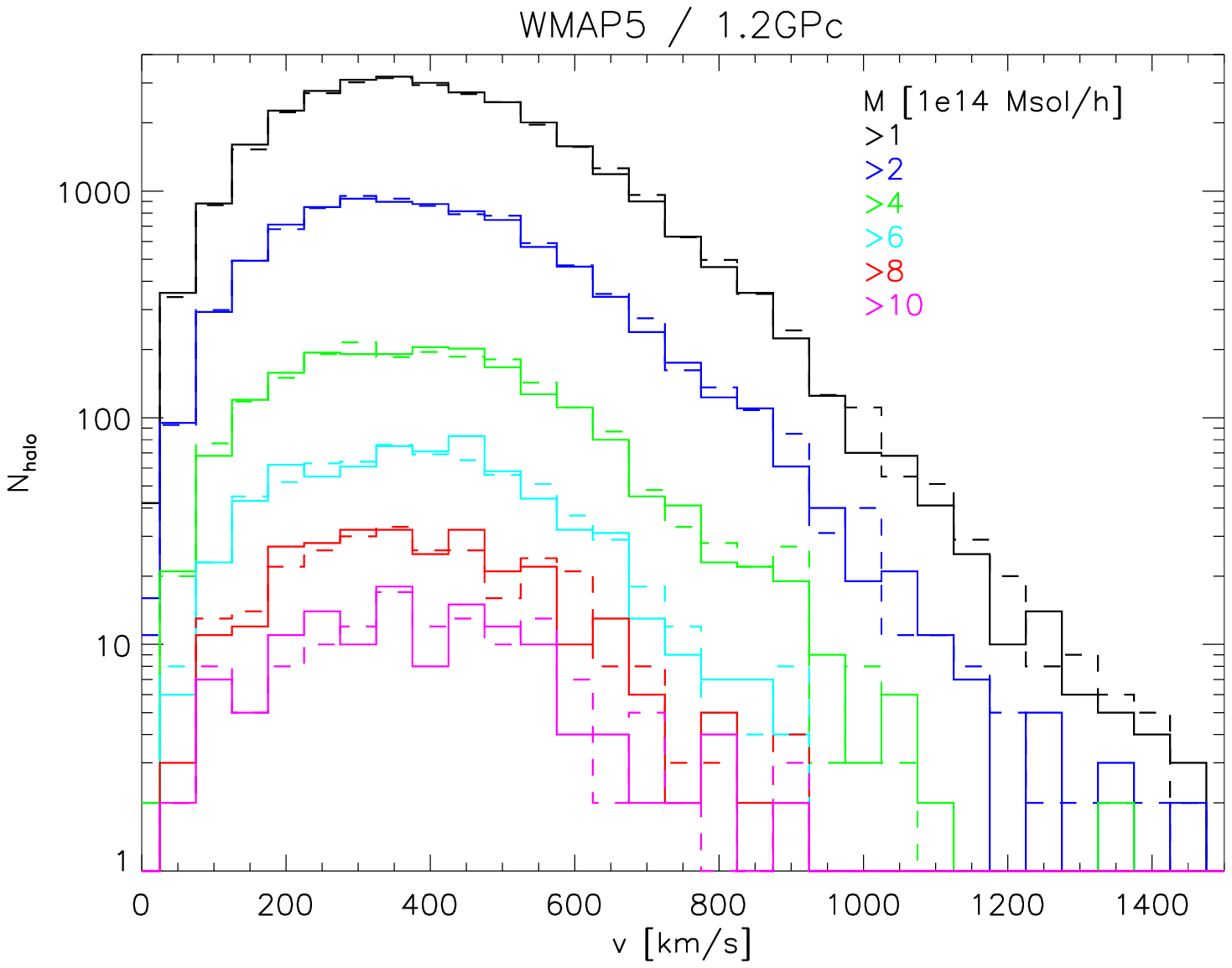}
\caption{Histogram of the peculiar motion of galaxy clusters from the large cosmological
box {\it WMAP5 / 1200} Comparing DM and csf simulation. Shown are the histograms for 
clusters above $10^{14}$, $2\times10^{14}$, $4\times10^{14}$, $6\times10^{14}$, 
$8\times10^{14}$ and $10^{15}$ $M_\odot/h$.}
\label{fig:halo_vel_hist_mass_gas}
\end{figure}

\begin{figure*}
\includegraphics[width=1.0\textwidth]{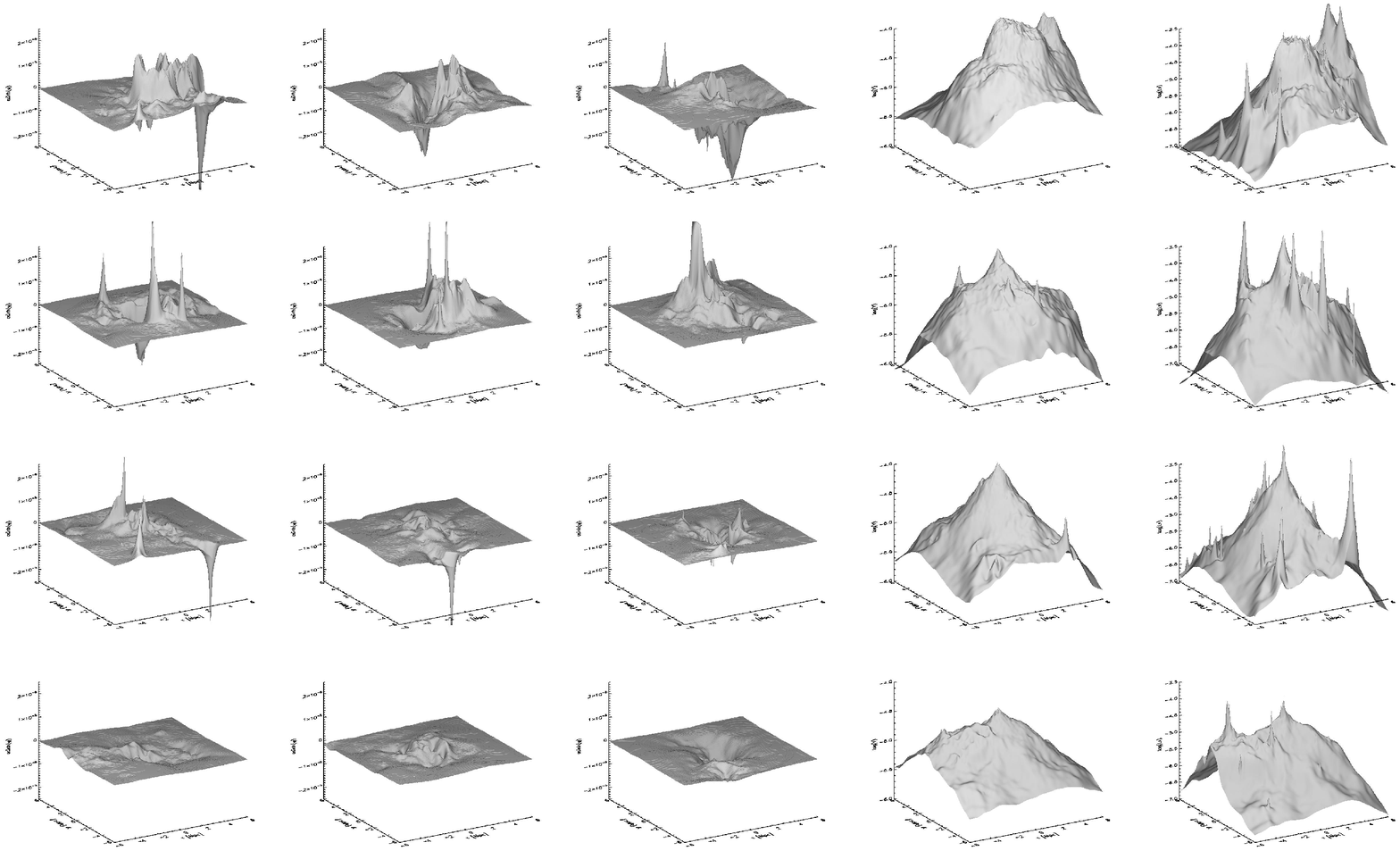}
\caption{Visualization of the kSZ, tSZ and X-ray SB for the 4 most massive clusters
within the medium size cosmological box {\it WMAP3 / 300} (4 rows). From left to right 
are the 3 different spacial projection directions (x,y,z) of the kSZ (shown is asinh(w)),
the tSZ (x projection, shown is log10(y)) and the X-ray surface brightness (x projection,
shown is log10(SB)). Displayed is a 1.5x1.5 Mpc region centered on the clusters.}
\label{fig:sz_collage}
\end{figure*}

\section{Spacial distribution of the observable signal}

To give an impression of the actual spacial distribution of the signal
we show in Figure \ref{fig:sz_collage} a visualize the maps of different 
observable, where the height in the 3D plot correspond to the strength of 
the signal. The region shown is a 1.5Mpc times 1.5Mpc region centered on 
the individual clusters. Shown is the kinetic (first three rows)
and thermal SZ effect (4th row) as well as the X-ray surface brightness 
of the four most massive clusters from the medium size cosmological box 
{\it WMAP3 / 300}. The internal motions of the different parts of the 
clusters can be clearly seen in the different minima's and maxima's of the maps.
Very sharp and isolated features are representing sub structures, which still
have a baryon content left which is not yet stripped by the cluster atmosphere.
In line with previous findings, only a very small number of the hundreds of resolved
substructures within the cluster have still some residual gas 
\citep[see also][]{2009MNRAS.399..497D}
which thereby contributes to the overall kSZ signal of the ICM. Due to the significant
internal bulk motions within the ICM also large scale features can be clearly seen.
Within the thermal SZ signal, the substructures with still host noticeable amount of gas
do not stick out so much compared to the ICM, demonstrating that the strong signals in 
the kinetic effect are dominated by the large, relative motion and are not due to a
larger comtonization parameter of the gas within such substructures. As most of the 
large scale structures within the cluster are approximately in pressure equilibrium,
the thermal SZ effect shows in comparison a much more regular structure. Also clear to
see that the X-ray signal, due to its density square dependence falls off much more
rapidly than the thermal SZ signal. 

\end{document}